\documentclass[preprint]{vgtc}                          

\ifpdf
  \pdfoutput=1\relax                   
  \usepackage{graphicx}                
  \DeclareGraphicsExtensions{.pdf,.png,.jpg,.jpeg} 
\else
  \usepackage{graphicx}                
  \DeclareGraphicsExtensions{.eps}     
\fi%

\usepackage{microtype}                 
\PassOptionsToPackage{warn}{textcomp}  
\usepackage{textcomp}                  
\usepackage{mathptmx}                  
\usepackage{times}                     
\usepackage{cite}                      
\usepackage{tabu}                      
\usepackage{booktabs}                  

\onlineid{0}

\vgtccategory{Research}

\usepackage[hyphens]{url} 
\vgtcinsertpkg
\hypersetup{final} 
\usepackage{setspace}
\usepackage{anyfontsize}
\usepackage{graphicx}
\usepackage[utf8]{inputenc}
\usepackage{xcolor}
\usepackage[clock]{ifsym}
\usepackage{xspace}

\newcommand{\versions}[2]{#2}

\graphicspath{{figures-medium/}} 

\newcommand{\myquote}[2]{{\setstretch{0.9}\begin{quote}\textit{{\fontsize{8.5pt}{10pt}\selectfont #1}}#2\end{quote}}}

\newcommand{\myinlinequote}[1]{``\textit{#1}''}

\definecolor{linkColor}{rgb}{0.1,0.1,0.8}
\newcommand{\imgsource}[2]{\texttt{{\color{linkColor}\href{#2}{#1}}}}

\newcommand{\permission}{\versions{\textit{Permission pending.}}{}}
\newcommand{\permissions}{\versions{\textit{Permissions pending.}}{}}

\newcommand{\vr}{{\sc vr}\xspace}
\newcommand{\ar}{{\sc ar}\xspace}

\newcommand{\mydots}{{\small [...]}\xspace}

\title{\versions{}{Towards }Immersive Humanitarian Visualization\versions{}{s}}

\author{Pierre Dragicevic\thanks{e-mail: pierre.dragicevic@inria.fr}\\ %
        \scriptsize Université de Bordeaux, CNRS, Inria, LaBRI}

\abstract{This paper introduces immersive humanitarian visualization as a promising research area in information visualization. Humanitarian visualizations are data visualizations designed to promote human welfare. This paper explains why immersive display technologies taken broadly (e.g, virtual reality, augmented reality, ambient displays and physical representations) open up a range of opportunities for humanitarian visualization. In particular, immersive displays offer ways to make remote and hidden human suffering more salient. They also offer ways to communicate quantitative facts together with qualitative information and visceral experiences, in order to provide a holistic understanding of humanitarian issues that could support more informed humanitarian decisions. But despite some promising preliminary work, immersive humanitarian visualization has not taken off as a research topic yet. The goal of this paper is to encourage, motivate, and inspire future research in this area.%
} 

\CCScatlist{
  \CCScatTwelve{Human-centered computing}{Visu\-al\-iza\-tion}{Visualization theory, concepts and paradigms};
  \CCScatTwelve{Human-centered computing}{Interaction paradigms}{Mixed / augmented reality}
}

\begin{document}


\firstsection{Introduction}
\label{sec:intro}

\maketitle

\myquote{If I am walking past a shallow pond and see a child drowning in it, I ought to wade in and pull the child out.
\mydots The fact that a person is physically near to us, so that we have personal contact with him, may make it more likely that we shall assist him, but this does not show that we ought to help him rather than another who happens to be further away. If we accept any principle of impartiality, universalizability, equality, or whatever, we cannot discriminate against someone merely because he is far away from us.}{\cite{singer1972morality}} 

This line of reasoning from the philosopher Peter Singer came to be widely known as ``the shallow pond argument''. According to this argument, if we can prevent something bad from happening at a low cost to ourselves, we should do it, whether the person who needs help is right in front of us or is an anonymous person in a remote country, and regardless of whether other people are also able to help. Accordingly, we should spend time and resources trying to alleviate tragedies elsewhere on the planet, for example by donating a small portion of our income to charities that can prevent deadly diseases \cite[Chap.~1]{singer2009life}. Although we might disagree with this moral imperative, it is a fact that (i) there exist countless potential actions that require small sacrifices from some people yet can bring immense relief to others, and (ii) increasing the frequency of those actions would benefit humanity. But there is a major psychological obstacle in doing so: not donating money to save an anonymous child far away feels very different from refusing to save the child drowning in the pond. It is still possible to relate to the plight of a distant child when viewing a documentary or a photo \cite{genevsky2013neural}, but when the child is just a data point in a dataset, the gap becomes enormous. Immersive humanitarian visualizations may help bridge this gap by informing people through quantitative facts, while making the plight of anonymous victims feel closer to the shallow pond situation.

I propose to refer to humanitarian visualizations as data visualizations or infographics designed to promote human welfare.\footnote{This is consistent with the Oxford dictionary definition of humanitarian as \textit{``concerned with or seeking to promote human welfare''}.} A quasi-synonym is the term ``anthropographics'', which was coined in information visualization \cite{boy2017showing} and defined as \myinlinequote{visualizations that represent data about people in a way that is intended to promote prosocial feelings (e.g., compassion or empathy) or prosocial behavior (e.g., donating or helping)} \cite{morais2020showing}. I introduce a new term for two reasons: first, my experience revealed that the term ``anthropographics'' causes lots of confusion, partly because it sounds a lot like ``anthropomorphic''. Yet the two have different meanings because not all visualizations designed to promote prosociality employ anthropomorphic figures, and vice versa \cite{morais2020showing}. In contrast, the word ``humanitarian'' is familiar and less likely to cause confusion. The second reason for introducing a new term is that in my view, promoting human welfare is more important than promoting prosocial feelings or behavior, and one does not necessarily imply the other. For example, a visualization that elicits deep empathy qualifies as an anthropographic, but empathy can cause distress and avoidance without necessarily leading to helping behavior \cite{bloom2017empathy}. An anthropographic can also inspire behavior that is well-intended but has a negligible or null impact on human welfare, as will be further discussed in \autoref{sec:effective}. Conversely, a visualization that helps a charity director allocate money across different health programs may not promote prosocial feelings or behavior (since all the money will be used to help people no matter what), but it can tremendously increase human welfare. Nevertheless, many anthropographics are humanitarian visualizations and vice versa, so most of the time it is acceptable to use the two terms interchangeably.

\begin{figure}[tb]
 \centering
 \includegraphics[width=0.47\columnwidth, trim=22mm 0 18mm 0, clip]{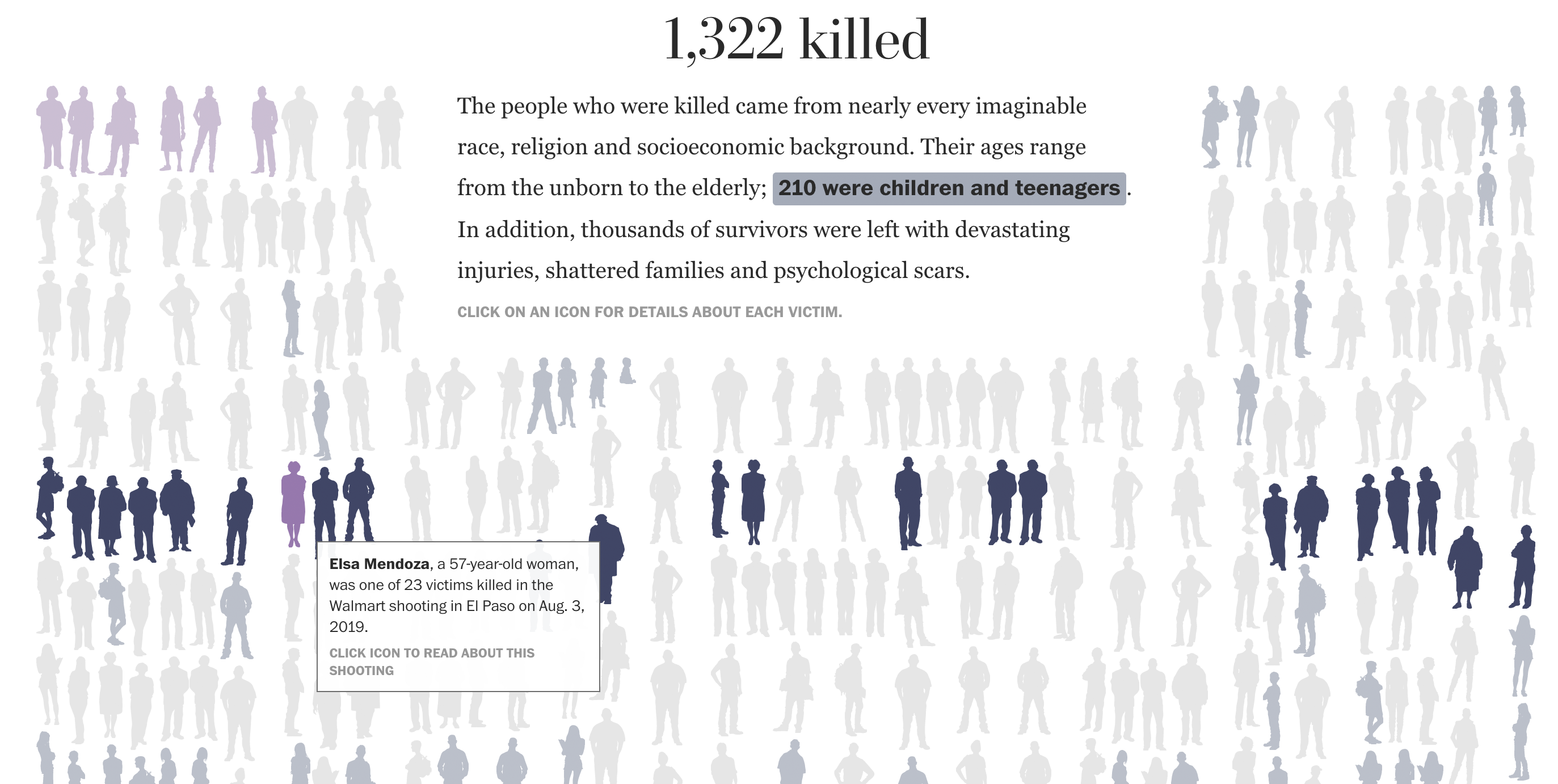}
 \includegraphics[width=0.51\columnwidth]{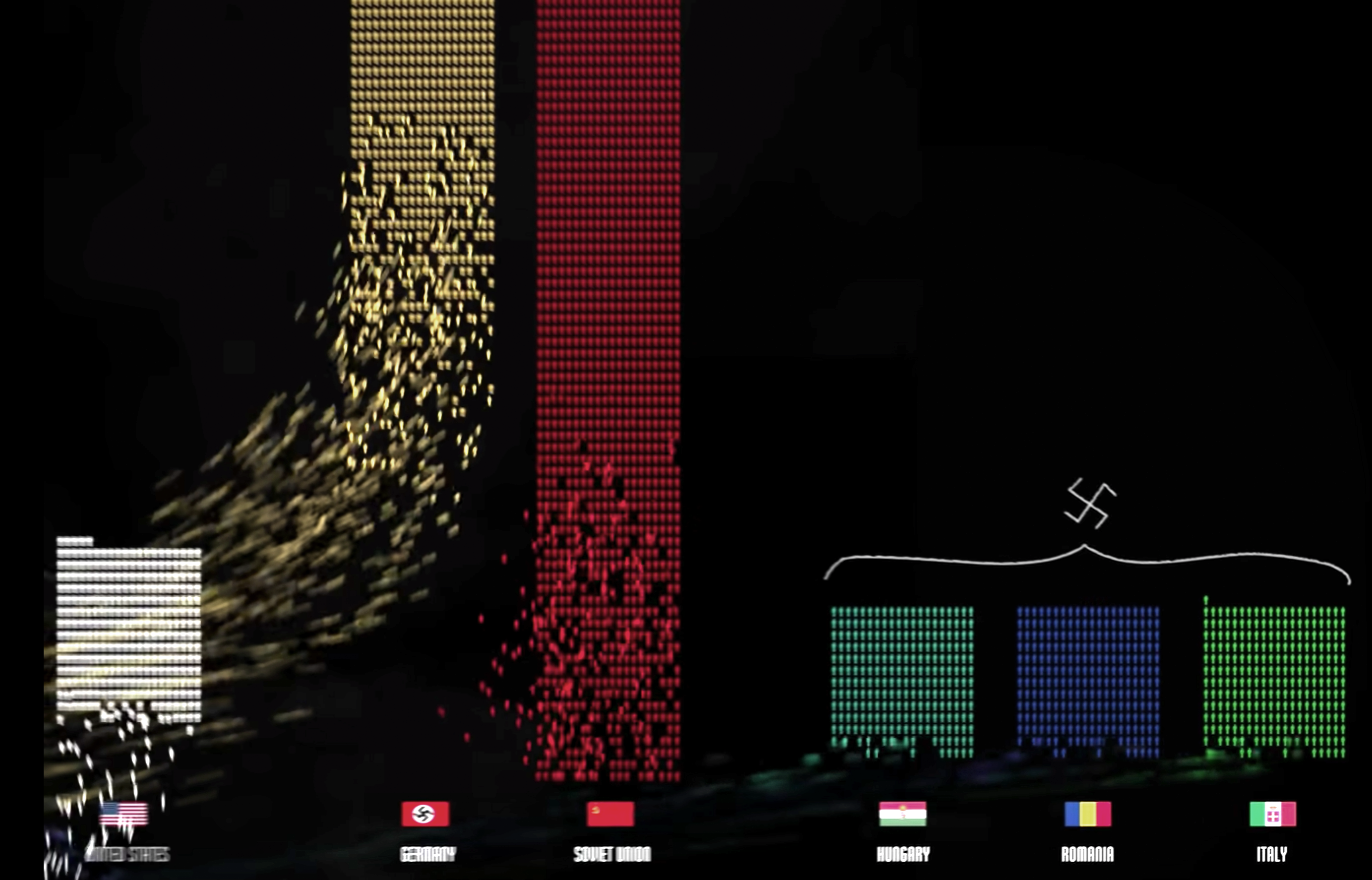}
\caption{Two humanitarian visualizations. \textit{Left:} interactive infographics by Bonnie Berkowitz and Chris Alcantara for the Washington Post, showing victims of mass shootings in the US. \textit{Right:} Documentary \textit{The Fallen of World War II} by Neil Halloran. Images from \imgsource{washingtonpost.com}{https://www.washingtonpost.com/graphics/2018/national/mass-shootings-in-america/} and \imgsource{youtube}{https://youtu.be/DwKPFT-RioU?t=433}. \permissions}
 \label{fig:examples-anthropographics}
 \vspace{-2mm}
\end{figure}

A lot has been written on how data visualizations can be useful for a variety of tasks (e.g., \cite[Chap.~1]{fekete2008value,card1999readings}). However, not much has been written on how it can promote human welfare, until recently, due to a surge of interest in the ethical, political and social dimensions of visualization~\cite{lupi2017data,correll2019ethical, VFSG}. In the past few years, data visualizations and infographics have been frequently used as means to raise the public awareness about humanitarian issues, especially in online journalism. A recent survey found more than 100 examples of such anthropographics \cite{morais2020showing, morais2020list}. \autoref{fig:examples-anthropographics} shows two examples. However, studies were subsequently conducted to test the intuitions behind common anthropographic design patterns, and they led to disappointing results -- a recent paper identified a total of 13 experiments reported in five studies, to which it added two extra experiments \cite{morais2021can}; Of the 15 experiments, most yielded inconclusive or negative results, and the few positive findings involved either very small effects or relied on weak statistical evidence requiring replications to be confirmed \cite{morais2021can}. The paper concluded:

\myquote{It remains possible that there exist alternative anthropographic design strategies that have a large effect on measures of affect, decision making, or behavior. However, the overall inconclusive results from the range of experiments conducted by researchers so far call for some skepticism. There is presently no clear evidence that if designers employ current anthropographic design strategies, this will have a clear and observable impact on people’s decisions and behavior.}{\cite{morais2021can}}

Immersive visualizations offer a rich and promising space to explore alternative design strategies. Like many communicative visualizations, most anthropographics have been designed for the web. Although this is a good way of maximizing reach, what is possible with web-based visualizations is limited by the capabilities of standard computing devices (desktop computers, smartphones). Researchers and designers should also explore solutions that go beyond regular displays and tap into the power of immersive displays, such as virtual reality (\vr) headsets. Actually, in parallel with the recent rise of anthropographics, there has been a big surge of interest in immersive displays for the purposes of visualizing information, which led to the emergence of a research area called immersive analytics \cite{marriott2018immersive}. However, with rare exceptions such as work from Ivanov et al. \cite{ivanov2018exploration, ivanov2019walk} which I will discuss soon, this stream of work has not been concerned with humanitarian applications.

Like previous work in immersive analytics \cite[Chap.~1]{marriott2018immersive}, I use the term ``immersive'' liberally, in a way that includes data visualizations on \vr displays, but also augmented-reality (\ar) visualizations, ambient visualizations, and physical visualizations (also called data physicalizations \cite{jansen2015opportunities}). Note that we often think of immersion as perceptual but it can also be purely cognitive (e.g., when reading a thrilling novel) \cite{isenberg2018immersive}. In this article, I will focus on cases where immersion is either perceptual, or both perceptual and cognitive. Because different immersive setups open up different possibilities, I will first discuss \vr systems where the viewer is fully immersed in a virtual world; I will then discuss cases where visualizations are integrated (or appear to be integrated) in the viewers' physical surroundings, which includes \ar, data physicalization, and ambient displays. I will then discuss the notion of effective altruism and explain why it is important to consider when designing immersive humanitarian visualizations. I will finally conclude by discussing related areas of research, ethical issues, and non-human animals.

Not all my discussions will involve data visualization as such -- I will discuss many cases where viewers are shown qualitative information and visual scenes without any data conveyed, because the most effective and interesting designs for immersive humanitarian visualizations are likely to combine elements of both worlds.

\section{Humanitarian Visualizations in Virtual Worlds}
\label{sec:VR}

When thinking about immersive humanitarian visualizations, the scene that first comes to mind is one where a viewer explores data about humanitarian issues while being immersed in a virtual environment that is separate from their physical surroundings.


\begin{figure}[tb]
 \centering
 \includegraphics[width=1\columnwidth]{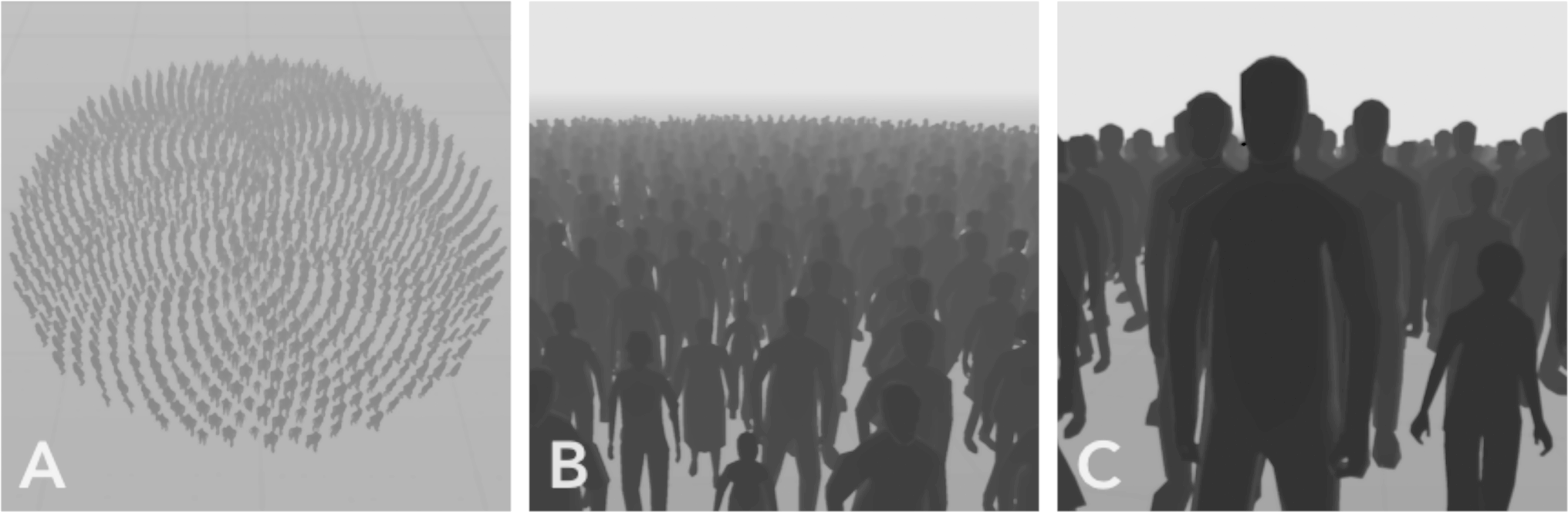}
\caption{Immersive humanitarian visualization by Ivanov and colleagues \cite{ivanov2018exploration, ivanov2019walk}. Each silhouette represents a person who died from a mass shooting in the US. Viewers can step back to get an overview of the dataset (A), or come closer to gather information about individual victims such as their age group or gender, which are encoded by the shape of the silhouette (B, C). Image from \cite{ivanov2019walk}. \permission}
 \label{fig:ivanov}
\end{figure}

\begin{figure}[tb]
 \centering
 \includegraphics[width=0.57\columnwidth]{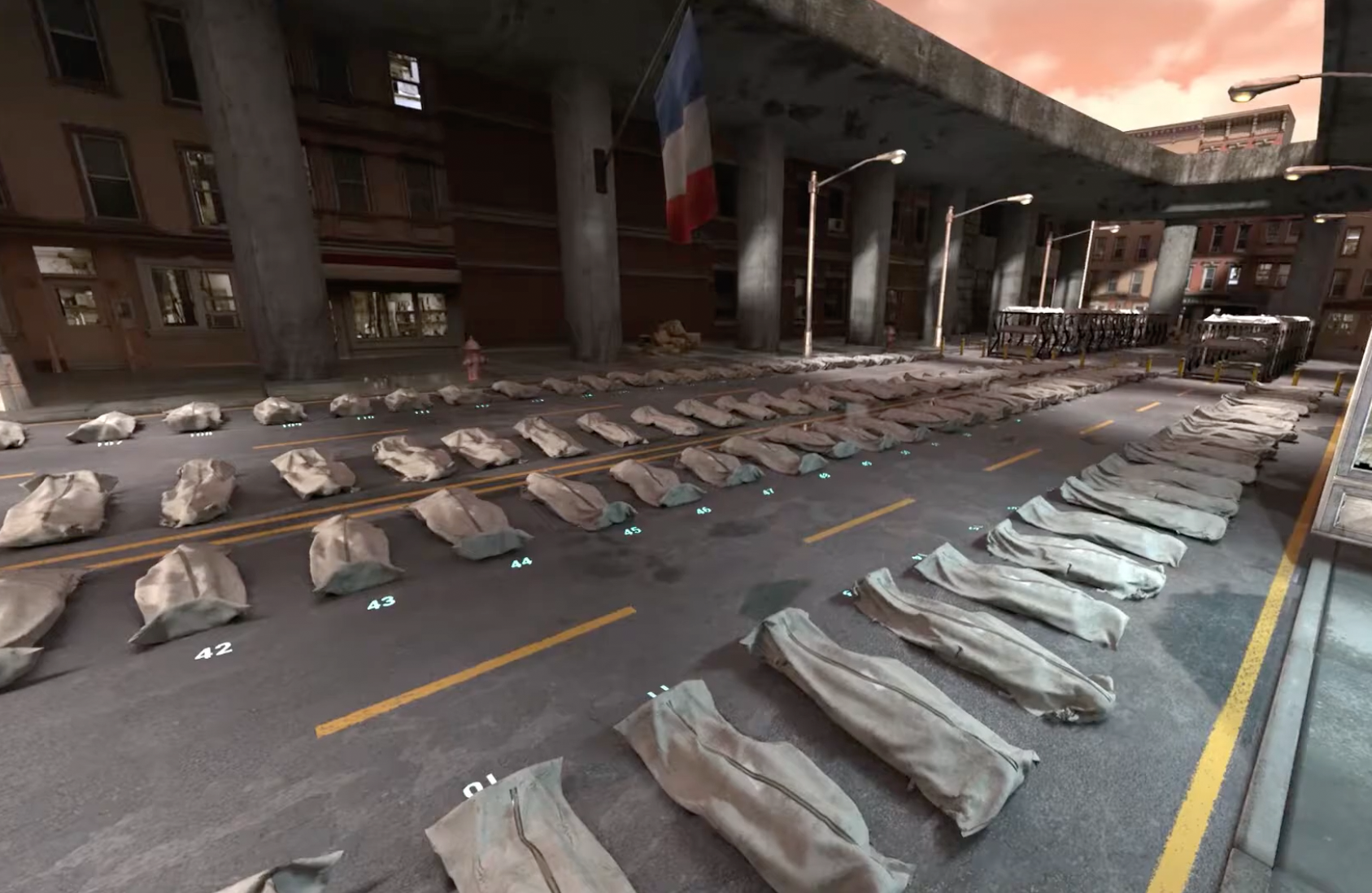}
\caption{Immersive humanitarian visualization by Ali Eslami titled \textit{DeathTolls Experience}. The viewer is shown data about three different mass casuality events (terrorist attacks in Europe, refugee deaths in the Mediterranean Sea, and the Syrian civil war). Each body represents one death. Image from \imgsource{youtube}{https://youtu.be/reReeH7CbEY?t=53}. \permission}
 \label{fig:eslami}
 \vspace{-2mm}
\end{figure}

The idea of using \vr to convey humanitarian data has been recently explored by Ivanov et al. \cite{ivanov2018exploration, ivanov2019walk, ivanov2019cg}, who designed a \vr visualization showing mass shooting casualties in the US (\autoref{fig:ivanov}). Each silhouette stands for a victim, and viewers see the entire dataset or get closer to individual people to learn more about them. They can also sort the data by gender, age, event, or US state, in which case avatars walk to form groups. This \vr system is among the very few existing contributions to immersive humanitarian visualization in the academic literature so far. Outside academia, there have been similar explorations by a \vr artist a few years back (\autoref{fig:eslami}). The artist designed a \vr world where viewers fly over hundreds of thousands of dead bodies representing victims of mass casualties. The content is delivered as a movie, so it is not interactive like the work by Ivanov et al. Also, no information is available about the victims, even though the rendering of the bodies is much more realistic. These two immersive visualizations are only preliminary explorations but they open up a range of possibilities, which I discuss next.

\subsection{Conveying More About Victims}

\begin{figure}[tb]
 \centering
 \includegraphics[width=0.80\columnwidth]{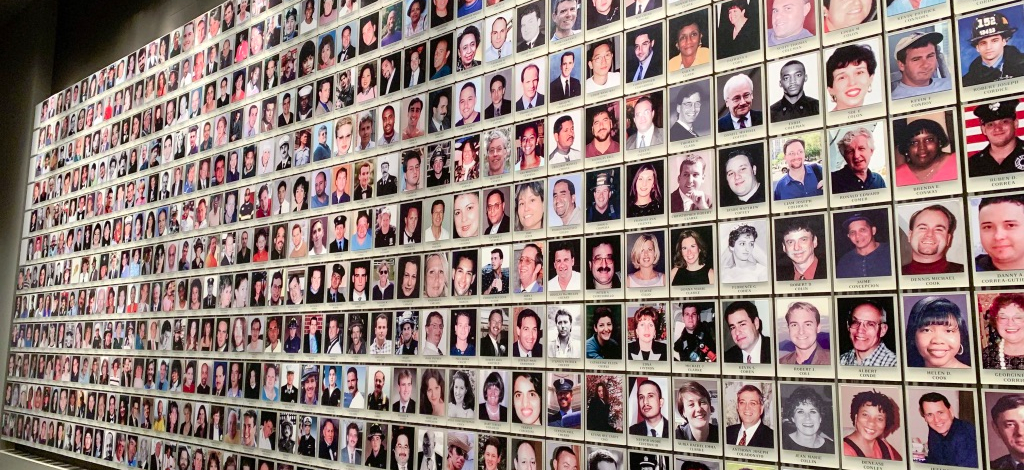}
 \vspace{-1mm}
\caption{Exhibit at the 9/11 Memorial Museum showing photos of 2,997 people killed during the 9/11 terrorist attack. Image from \imgsource{www.hawkeyenews.net}{https://www.hawkeyenews.net/features/2019/05/06/9-11-memorial/}. \permission}
 \label{fig:9-11-memorial}
\end{figure}

In the two visualizations we have just seen, the models that stand for dead people look visually very similar, which may contribute to deindividualizing them. It would be useful to explore the effects of conveying richer information about individual victims, both quantitative and qualitative.
A simple example of rich qualitative information would be realistic depictions of the victims.\footnote{When discussing hypothetical designs, I will not consider how and whether they can be implemented with currently available technology.} Ivanov et. al mention having tried more realistic 3D models than their silhouettes, but they gave up because the imperfections in the models and the fact that they were all clones of the same female and male templates yielded uncanny valley effects \cite{ivanov2019walk}. Using highly realistic depictions of the victims themselves will likely eliminate uncanny valley effects, though it might still make viewers feel uneasy; Although realistic depictions of diseased individuals are commonplace in many cultures (e.g., photo portraits in obituaries, portraits and sculptures celebrating famous people, photos and sculptures in memorials -- see \autoref{fig:9-11-memorial}), being immersed among realistic and potentially animated replicates of diseased individuals could be disturbing. Regardless, potential adverse effects on viewers will be important to consider, as well as the preservation of the privacy and humanity of the victims; I will discuss these and other ethical issues in \autoref{sec:ethics}.

\subsection{Conveying Personal Experiences}
\label{sec:personal-experiences}

In addition to conveying factual personal information, immersive humanitarian visualizations could convey information about each victim's personal story, how they died, or the suffering they went through; Or if they are still alive, the hardship they are currently going through, or they could be facing if they do not get help.

For conveying personal experiences, we can draw from the rich body of work on using \vr to promote empathy and perspective-taking. \vr has been touted as the ``ultimate empathy machine'' \cite{milk2015how} due to its ability to realistically reproduce experiences from other people, be they real or hypothetical. For instance, a variety of immersive documentaries have been made that use footage from stereoscopic or 360-degree cameras; A 2018 survey found more than 80 of them on the topic of war and conflict alone, and more than 40 on migration and displacement \cite{bevan2018mediography,bevan2018history}. To be sure, regular documentaries and movies (and before them, books and oral stories) have been long used to evoke empathy. But \vr displays might bring an additional dimension due to their increased level of perceptual immersion and the interaction possibilities they offer. \vr designers and researchers have indeed experimented with many perspective-taking concepts involving for example adopting a body of a different shape \cite{serino2016virtual}, skin color \cite{peck2013putting} or gender \cite{bolt2021effects}, or swapping bodies in real time \cite{verge2014using,caters2016vr}. Meanwhile, a genre of indie video game has recently emerged whose purpose is to convey human experiences such as being a woman refugee from Syria \cite{pixel2018bury}, managing a refugee camp \cite{gyroscoping2022building}, or parenting a toddler with terminal cancer \cite{numinous2016that}.\footnote{Thanks to Wesley Willett for pointing these out to me.} 

A lot more could become possible if anyone was able to easily capture and share immersive footage of their intimate lives for educational or fundraising purposes. Charities have long had sponsorship programs where potential donors can see photos and basic information about potential recipients (usually children). Similarly, the GiveDirectly organization is in the process of launching a program where individual donors are paired with specific recipients and will see whom their money goes to \cite{faye2021cash}. Insights about the potential usefulness of immersive footage over simple photos can be gained by looking at the opposite of saving lives remotely: killing remotely. Research on drone warfare indeed suggests that many drone operators find it extremely challenging to kill a person after having observed their lives for days, weeks or months, and the longer the observation and the higher the resolution of the video feed, the harder it is for them \cite{phelps2021killing}. Turning this finding on its head, perceptually connecting with the lives of remote people could promote helping behavior, and the longer the time spent and the higher the level of visual detail (and perhaps immersion), the larger the effect may be. 


%

\subsection{Mixing Quantitative Facts and Experiences}
\label{sec:unfolding-tragedies}

For researchers and designers, a key question will be how to use \vr to convey factual quantitative information about populations and individuals, while at the same time sharing personal experiences the way videos and interactive stories do. One way to do this could be to apply the ``near and far'' pattern from data journalism \cite{harris2015connecting} and allow viewers to seamlessly transition from data to individual experiences. For example, if the focus is on conveying past tragedies like mass shootings, a viewer could explore a unit visualization like the one in Figure 3 and ``possess'' the person of their choice to relive their personal story, possibly reconstructed from historical records, notes from personal diaries, or witness testimonies.


There is a clear educational benefit in learning about past tragedies, but there is also great value in conveying information about ongoing and future (hypothetical) tragedies, if only because viewers may be able to help prevent or minimize such tragedies. For example, one could imagine a \vr reporting service offering rich and up-to-date information about an ongoing war. Viewers could first view a map of the invaded country showing the location of the troops, then zoom into cities to view data about how inhabitants are affected (bombings, food or power shortages,...), and finally dive in the streets and houses to experience people's lives by viewing recent or real-time video footage. Although such an application could promote undesirable sentiments like voyeurism, it could also be one of the very few ways the reality of war can be fully conveyed to people who wish to get a deep understanding of the situation, on an intellectual and on a visceral level. Because such an application adds an experiential dimension to factual information, it is not absurd to think that it could also help decision makers like army officers and politicians make more informed and more humanitarian decisions.

\begin{figure}[tb]
 \centering
 \includegraphics[width=0.95\columnwidth]{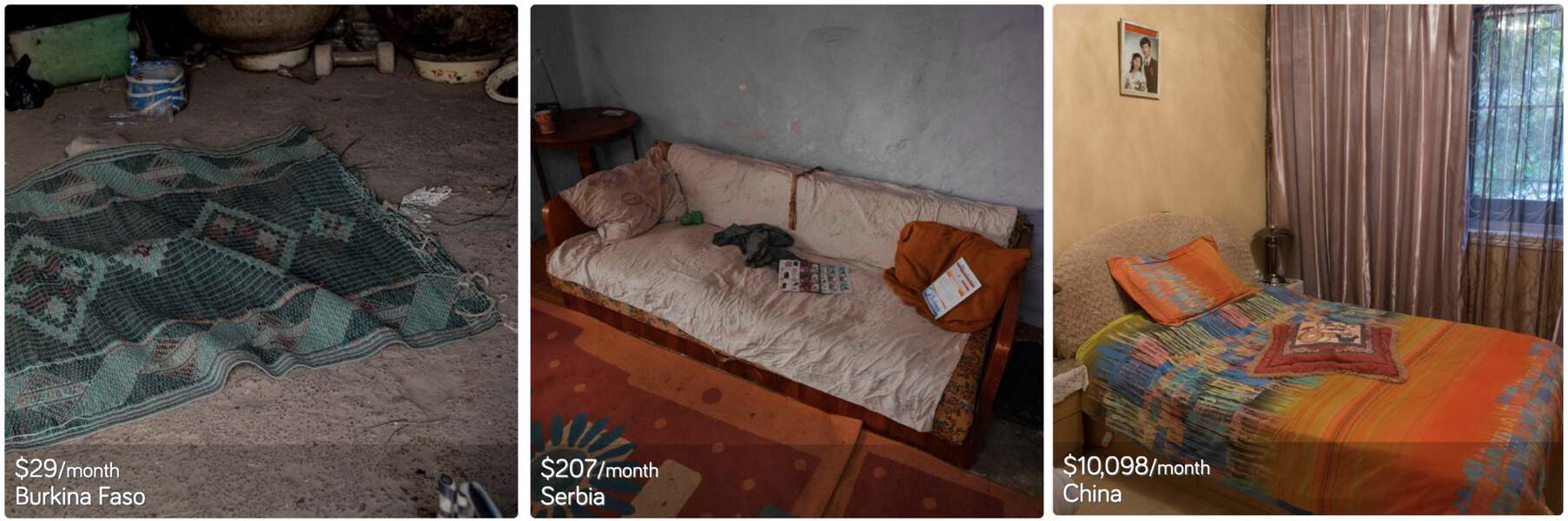}
\caption{Dollar street. Photos of beds of people with different incomes and different living places. \imgsource{www.gapminder.org}{https://www.gapminder.org/dollar-street?}. Images CC BY 4.0.}
 \label{fig:dollar-street}
 \vspace{-2mm}
\end{figure}

Besides informing people about humanitarian disasters, immersive humanitarian visualizations can be useful for communicating about global poverty issues, for example by conveying both quantitative and qualitative information about the quality of life of people in disadvantaged regions of the planet. Although it is not immersive, a great source of inspiration is the Dollar Street web app (\autoref{fig:dollar-street}), which lets viewers explore a database of 30,000 photos of 264 families with differing incomes across 50 countries \cite{gapminder2016dollar}.\footnote{The 1994 book ``Material World'' \cite{menzel1994material} explored a similar concept by showing photos of families with all their possessions across 50+ countries.} Viewers can filter by income or location and compare photos of people’s beds, their toilets, but also things like their teeth, their children’s toys, or how they wash their hands. This application illustrates an original way of combining quantitative and qualitative information: it shows purely qualitative content (photos), but viewers can filter and organize this content based on quantitative data. A variation of this principle is demonstrated by the work by Concannon et al. \cite{concannon2020brooke}, who made multiple versions of a video telling the story of a fictional care leaver in England (a young adult who is transitioning from foster care to an independent life). Each version tries to imagine the life of the protagonist depending on where they live in England, based on real data about local policies for supporting care leavers -- the worse the policy, the more difficult the life of the protagonist.

\begin{figure}[tb]
 \centering
 \includegraphics[width=0.50\columnwidth]{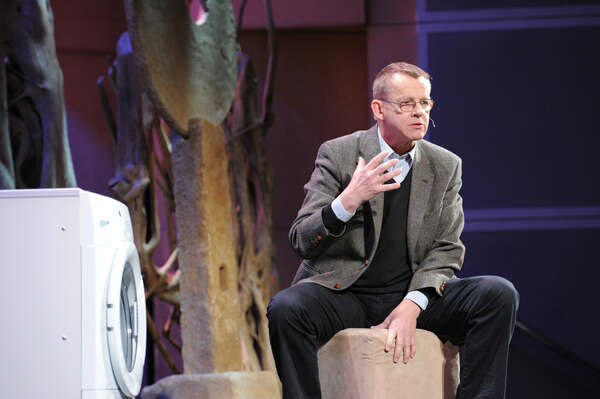}
 \includegraphics[width=0.46\columnwidth]{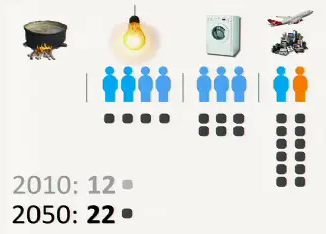}
\caption{Excerpts form Hans Rosling's TED talk ``The Magic Washing Machine''. Images from \imgsource{www.ted.com}{https://www.ted.com/talks/hans_rosling_the_magic_washing_machine}. \permission}
 \label{fig:rosling}
\end{figure}

\begin{figure}[tb]
 \centering
 \includegraphics[width=1\columnwidth]{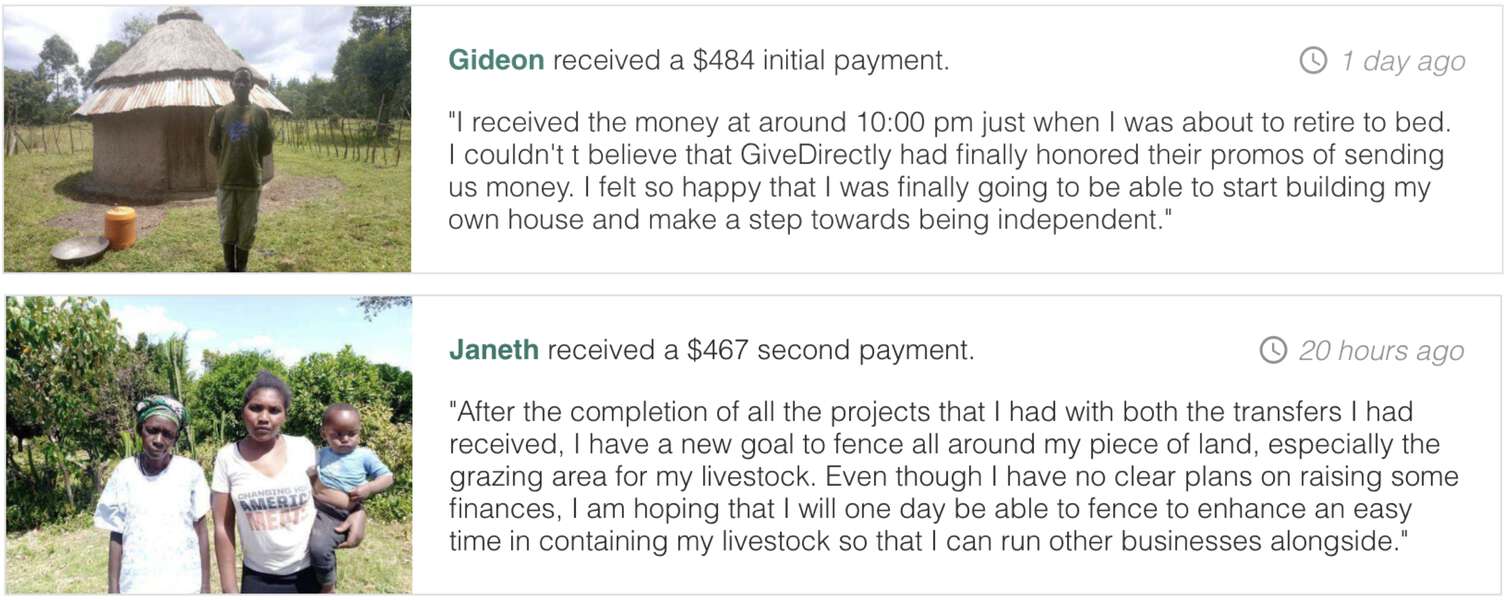}
\caption{Excerpts from GDLive web page from the GiveDirectly organization (\imgsource{live.givedirectly.org}{https://live.givedirectly.org}) that posts information and updates about cash transfers, including what selected recipients will do with the money, their reaction when they get the money, and how their lives was changed afterwards. \permission}
\vspace{-2mm}
 \label{fig:gdlive}
\end{figure}

\subsection{Tapping Into Positive Framings}
\label{sec:positive}

So far I focused on how immersive humanitarian visualizations can be used to communicate negative situations, but they can also be used to communicate positive situations. One non-immersive but again inspiring example is the talk from Hans Rosling ``The Magic Washing Machine'' (\autoref{fig:rosling}) \cite{rosling2010magic}. He first tells a story that gives a powerful account of how life-changing washing machines are, and then goes through data about how many people in the world have access to them, and how this is likely to change with economic growth. In contrast to the approach of viewing data first and then zooming into individuals (\autoref{sec:unfolding-tragedies}), here the personal story comes first and the data follows. As another example of positive framing, the GiveDirectly organization has a web page with a news feed where people from rural regions of East Africa share their reactions where they just got a cash transfer, or explain what they will do with the money, or how a previous cash transfer changed their lives (\autoref{fig:gdlive}). They will for example report that they were able to pay their children's schooling debts, that they can now afford healthier and more frequent meals, that they could buy a cow to get milk everyday, or that they were able to repair their house. Likewise, immersive humanitarian visualizations could focus on the positive sides of charitable donations, by conveying both quantitative and qualitative information about how past donations improved people's lives, or how hypothetical donations can do the same. We have already seen that connecting potential donors with the lives of potential recipients could motivate donations. Going further and connecting donors with the lives of their recipients \textit{after} they have made a donation might have a rewarding effect that motivates subsequent donations.

\section{\scalebox{.97}[1.0]{\hspace{-1mm}Humanitarian Visualizations in the Physical World}}
\label{sec:AR}

\myquote{Jeff McMahan: \mydots The psychology is quite different when you write a check and when you actually go out into the world and help people in a face-to-face sort of way.\vspace{2mm}\\Sam Harris: \mydots It seems to me that we want to close the gulf in a variety of ways, because we want the distant suffering of people we will never meet to be more salient the more we actually can do something about it. In a world where you can do nothing about it, then maybe there's an argument for ignorance being bliss. But in a world where our own economic activity, at least, affects people all over the globe, and we are implicated in their suffering in the way we marshal our resources, it seems to me, we want to make it more salient. And then when we jump into some virtuous side of the machine, and do more and more good simply by cutting a check or using our resources, or time and attention in various ways, we want to make it more and more analogous to the very good feels you would get actually being a face-to-face hero. And so anything that would allow us to do that, I think would be good. And I think we want it to be as rewarding as it can be, and as it would be, if we did it up close and personal.}{\footnote{Excerpt from audio conversation \cite[starts at 1:04:00]{harris2021can}. The audio transcript has been slightly edited for legibility.}}

This interview excerpt played a key role in inspiring this section. Sam Harris' vision does not involve people occasionally immersing themselves in \vr stories and data, as they would read newspaper articles or fundraising calls about humanitarian crises. Instead, his vision involves changing the way people relate to their daily lives, and how they perceive the world and make decisions as they go about their day. I think embedding humanitarian visualizations in our physical environments can help us implement this vision.

There are two major ways in which humanitarian visualizations can be embedded in our physical environments \cite{willett2016embedded}: either by using augmented reality displays to give the illusion that they are in our environment, or by actually embedding visualizations in our environment, for example as data physicalizations \cite{jansen2015opportunities} or ambient displays \cite{moere2007towards, elmqvist2013ubiquitous}. Although all such setups provide a very different kind of immersion than virtual reality displays, they are still immersive in the sense that the world is immersive: the physical environment that surrounds us often yields a richer perceptual experience and richer opportunities for action than applications running on desktop computers or smartphones \cite{thomas2018situated}.

\begin{figure}[tb]
 \centering
 \includegraphics[width=0.555\columnwidth]{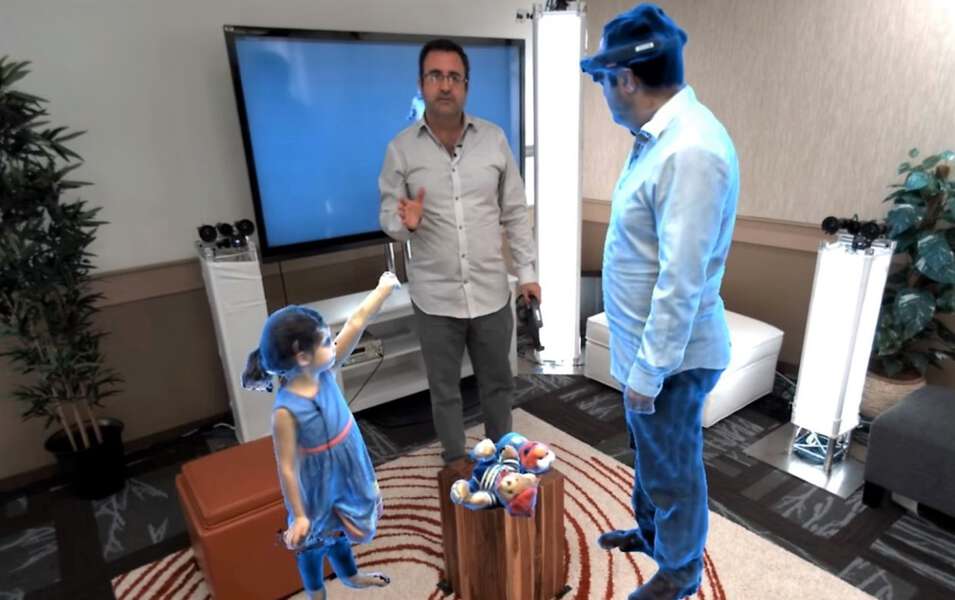}
 \includegraphics[width=0.435\columnwidth]{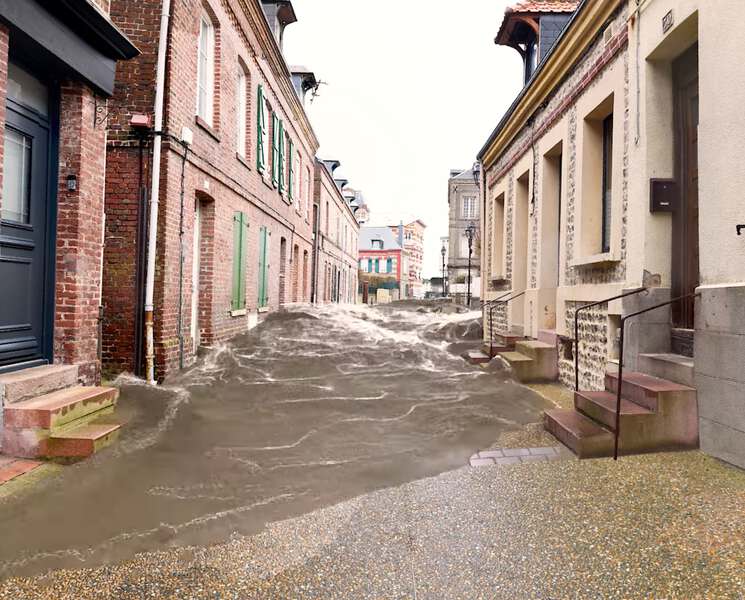}
\caption{\textit{Left:} Holoportation, a technology that uses 3D capture and \ar to make remote people appear as if they were near the viewer \cite{orts2016holoportation}. \textit{Right:} simulation of sea rise in the city of Etretat, France. Images from \imgsource{www.geekwire.com}{https://www.geekwire.com/2016/microsoft-research-star-wars-holoportation-hololens/} and \imgsource{youtube}{https://www.youtube.com/watch?v=_YZYGnWMzFc}. \permissions}
 \label{fig:AR_relocation}
 \vspace{-2mm}
\end{figure}

\subsection{Relocation with Augmented Reality}

\ar displays offer unprecedented opportunities due to their ability to create illusions of objects and people around us, which includes relocating distant objects or people in the viewer's vicinity. Technologies have been already demonstrated that can capture in real-time high-quality 3D models of people, objects and environments, and display them in the physical environment of remote viewers (\autoref{fig:AR_relocation}-left)  \cite{orts2016holoportation}. Such systems open up the possibility of bringing the lives of distant suffering people closer to our own. Immersive video scenes of personal lives could take on a whole new meaning, as \ar could allow viewers to insert them in their own physical spaces instead of viewing them on a screen or having a \vr display transport them elsewhere. Individuals on the other side of the world could temporarily become our neighbors or roommates. Besides people, we can imagine entire rooms, homes, or even large-scale outdoor scenes that are temporarily relocated near us. This can help make humanitarian issues more salient or more memorable -- for example, if a person walks in a refugee camp that has been temporarily relocated in their backyard, they may create a mental association and remember the refugees each time they see (or even think about) their backyard. In contrast, \vr can subjectively transport viewers in distant places, but once the viewers are back, the event is remote again. Augmented reality can also tap into the affective relations we have with our physical environment -- for example, if a remote city is being bombed or hit by natural disaster, we could see bombs fall on our own city\footnote{The very evening after I wrote this, a realistic mock-up video showing Paris being hit by bombs was released on the Ukrainian parliament's Twitter account: \imgsource{twitter link}{https://twitter.com/ua_parliament/status/1502402021386858504} \imgsource{(archive)}{https://web.archive.org/web/20220315231428/https://twitter.com/ua_parliament/status/1502402021386858504}} or water flood our own streets (\autoref{fig:AR_relocation}-right). As another example, people protesting a totalitarian regime in a remote country could be protesting in our own streets at the same time.

With future wearable \ar technology, it is even possible to imagine that some viewers will choose to have such scenes visible for extended periods of time while they go about their day, or shown intermittently, as reminders of the suffering of remote people. In this case, the lives of the suffering people would become even more deeply intertwined with their own. Granted, it can be hard to imagine that many people will willingly subject themselves to experiences akin to being haunted by ghosts or hallucinations. However, people already engage in unpleasant activities for future personal gain such as losing weight or becoming healthier, so it is not unreasonable to think that some may choose to do so for altruistic purposes, perhaps because they find their willpower lacking and need a commitment device \cite{bryan2010commitment} to do good, or do good more frequently.

Instead of only conveying negatives, such \ar systems could also use positive framings, as I already suggested in \autoref{sec:positive}. For example they could show  improvements in people's lives brought about by charitable donations, or by the viewer's own contribution. In the context of a donor/recipient pairing program (see \autoref{sec:personal-experiences}), with future \ar technology, it is possible to imagine that a donor occasionally meets a past recipient on the street and chats with them: a long-distance cash transfer may suddenly feel like helping out an acquaintance in a small village.

\begin{figure}[tb]
 \centering
 \includegraphics[width=0.58\columnwidth]{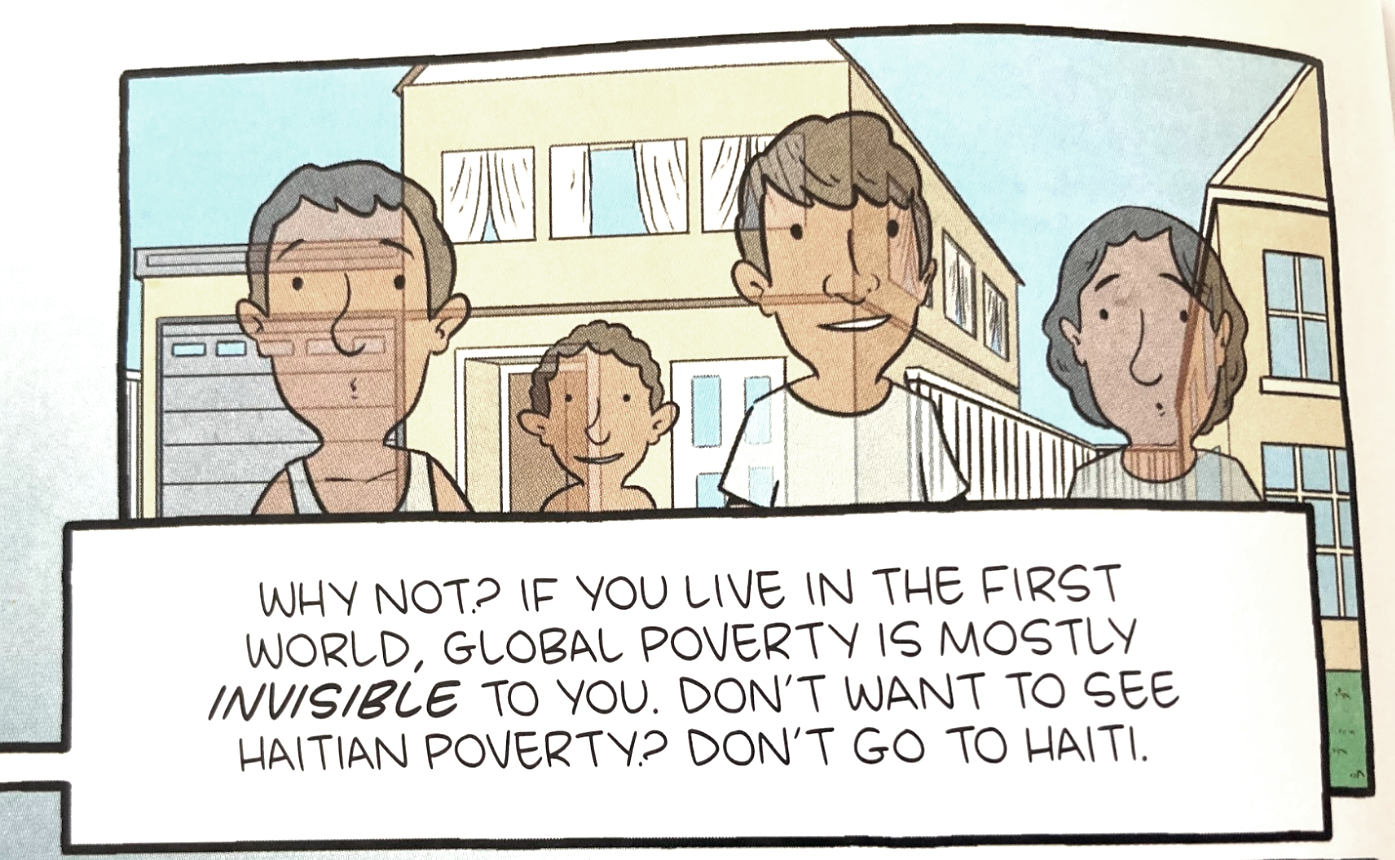}
 \includegraphics[width=0.40\columnwidth]{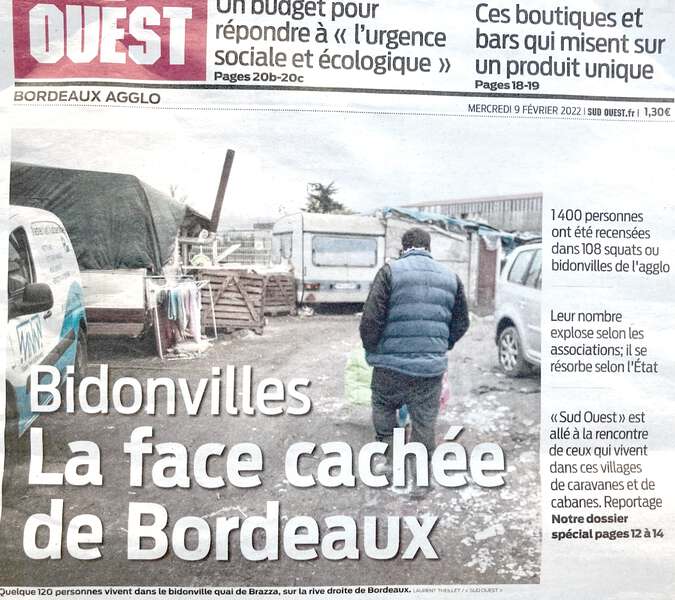}
\caption{The invisibility metaphor used to refer to both distant and local poverty. \textit{Left:} excerpt from the graphic novel Open Borders \cite{caplan2019open}. \textit{Right:} first page of the newspaper Sud-Ouest titled ``Slums -- the hidden side of Bordeaux''. \permissions}
\vspace{-2mm}
 \label{fig:invisible}
\end{figure}

There are several possible variations around the idea of relocating distant people closer to potential helpers. First, the people who are virtually relocated need not be very far away -- even the richest cities in the world have inhabitants whose suffering goes largely ignored, such as hospitalized patients, prison inmates, or people in poor housing conditions (\autoref{fig:invisible}). Second, scenes featuring relocated people and places can be reconstructions based on data and documentary evidence instead of real scenes; This would preserve people's privacy and would allow to convey issues such as domestic violence, for which the use of real scenes is not an option.

\begin{figure}[tb]
 \centering
 \includegraphics[width=0.52\columnwidth]{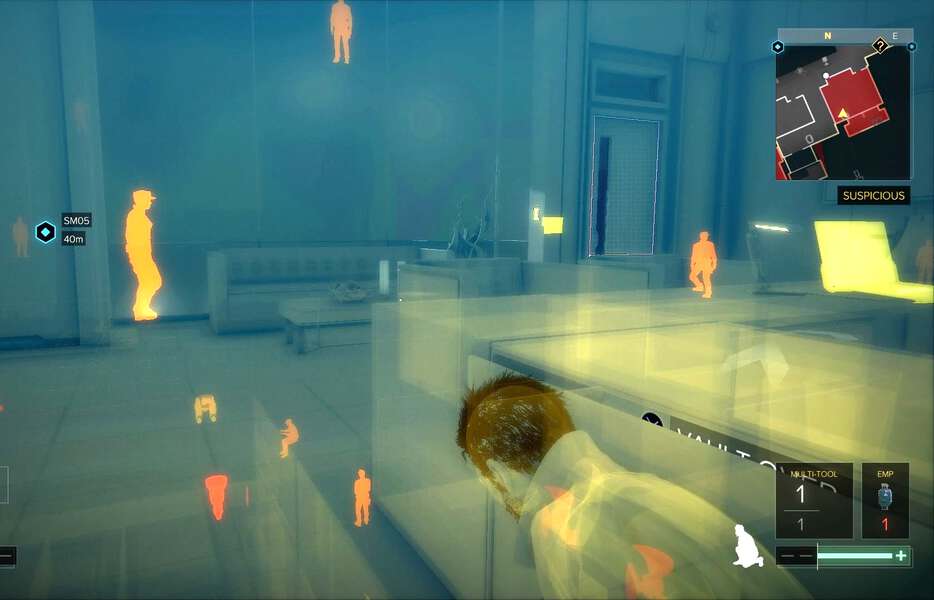}
 \includegraphics[width=0.47\columnwidth]{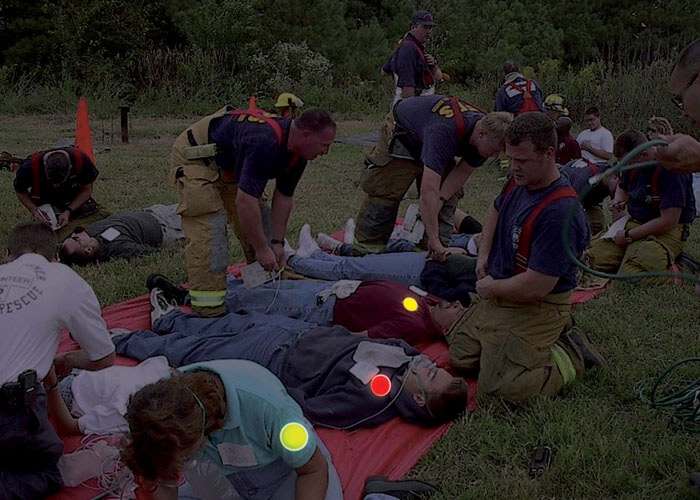}
\caption{Ways of revealing hidden people or hidden information about people. \textit{Left:} In the video game Deus Ex: Mankind Divided, the protagonist can use an electronic implant that allows him to see people through walls. \textit{Right:} Triage lights used in scenes of accidents to prioritize medical care (red = require immediate attention, yellow = stable for the moment). Images from \imgsource{www.mobygames.com}{https://www.mobygames.com/game/playstation-4/deus-ex-mankind-divided/screenshots/gameShotId,881189/} and \imgsource{www.cyalume.eu}{https://www.cyalume.eu/en/baton-lumineux-cyalume-chemlight-snaplight/batons-et-marqueurs-lumineux/marqueur-lumineux-circulaire-adhesif-lightshape/triage-medical-avec-cercles-lumineux-cyalume-lightshape/}. \permissions}
 \label{fig:deus-triage}

\end{figure}

\subsection{Enhanced Vision with Augmented Reality}

Other metaphors than relocation can be used to make hidden people visible, including x-ray vision, or the ability to see behind walls. This metaphor is already used in video games (\autoref{fig:deus-triage}-left) and in some \ar visualization prototypes \cite{willett2021superpowers}. Besides showing where people are situated, \ar systems could use rich visual encodings to convey their physical and mental state, including whether they experience pain or distress, and the intensity of those negative sensations. Such a tool could greatly help professionals such as hospital medical personnel, hospice nurses, and first responders reach faster and more informed decisions. Inspired by current practices in medical triage (\autoref{fig:deus-triage}-right), Willett et al. \cite{willett2016embedded} similarly discuss a speculative scenario where drones are deployed in a disaster area to locate survivors, assess their level of criticality, and send this information back to human rescuers using colored lights. The authors further mention that 
\myinlinequote{seen together (for example from the air), the complete set of illuminated drones could also serve as an aggregate visualization, highlighting important areas and providing an overview of the environment.} \cite{willett2016embedded}. A visualization relying on \ar headsets instead could show richer information about individuals as well as information about people hidden by buildings and walls. Beyond professional applications, similar tools could be used by the general public as reminders of otherwise invisible suffering, be it nearby (e.g., a hospital) or far away (e.g., a war in a remote country). Even if the directional information is not actionable in this case, seeing it could make the suffering more concrete, just as praying towards a remote holy place makes the place more tangible in one's mind.

\begin{figure}[tb]
 \centering
 \includegraphics[width=0.45\columnwidth]{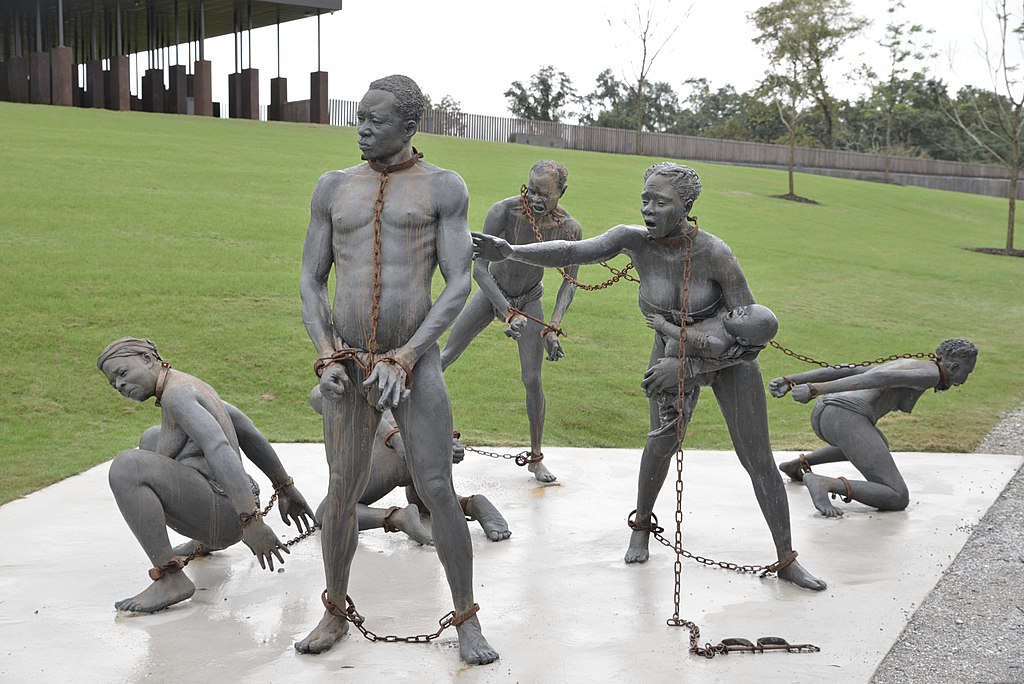}
 \includegraphics[width=0.54\columnwidth]{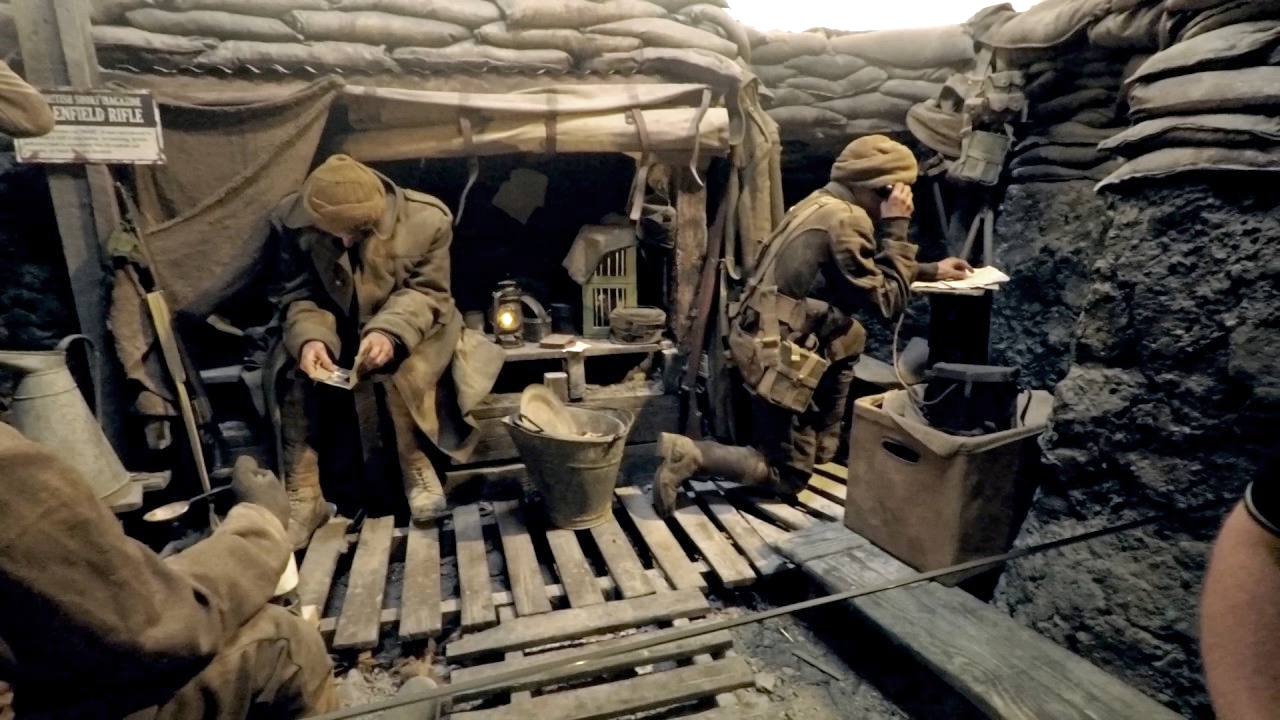}
\caption{Physical reproductions of humanitarian tragedies. \textit{Left:} The National Memorial for Peace and Justice in Montgomery, Alabama -- a sculpture commemorating the victims of racial terror lynching in the US. \textit{Right:} Life-size diorama at the The Great War Exhibition, Wellington, NZ. Images from \imgsource{joyofmuseums.com}{https://joyofmuseums.com/museums/united-states-of-america/montgomery-alabama-museums/memorial-for-peace-and-justice/} and \imgsource{Tim Simpson}{https://www.youtube.com/watch?v=npAGzEBMlGw}. \permissions}
\vspace{-2mm}
 \label{fig:physmodels}
\end{figure}

\subsection{Sculptures and Data Sculptures}

I discussed cases where physical spaces can be augmented by optical illusions, using \ar technology, but they can also be augmented physically. Although many of the examples I previously discussed would be impossible to implement in this manner, this is still a rich and promising area to explore. Augmenting our environment with physical objects conveying human suffering is already common practice -- public places and museums display statues and sculptures as memorials of previous humanitarian tragedies (\autoref{fig:physmodels}-left). Such objects are much more deeply integrated in the physical environment than current augmented-reality models because they are visually realistic, they can be touched, they are always present, and they do not need special equipment to be seen. On the other hand, they are static and cannot be easily replicated and transported (e.g., in somebody's home) unless they are made much smaller. Similarly, it is possible to re-create reasonably large scenes (\autoref{fig:physmodels}-right)\footnote{For another striking example see \imgsource{www.rferl.org}{https://www.rferl.org/a/russia-creates-worlds-biggest-world-war-2-diorama/30169029.html} and \imgsource{youtube}{https://www.youtube.com/watch?v=as1KQ_nseRk} video.}, but physical models are inadequate for capturing large-scale human tragedies that are unfolding like city bombings or natural disasters.

\begin{figure}[tb]
 \centering
 \includegraphics[width=0.58\columnwidth]{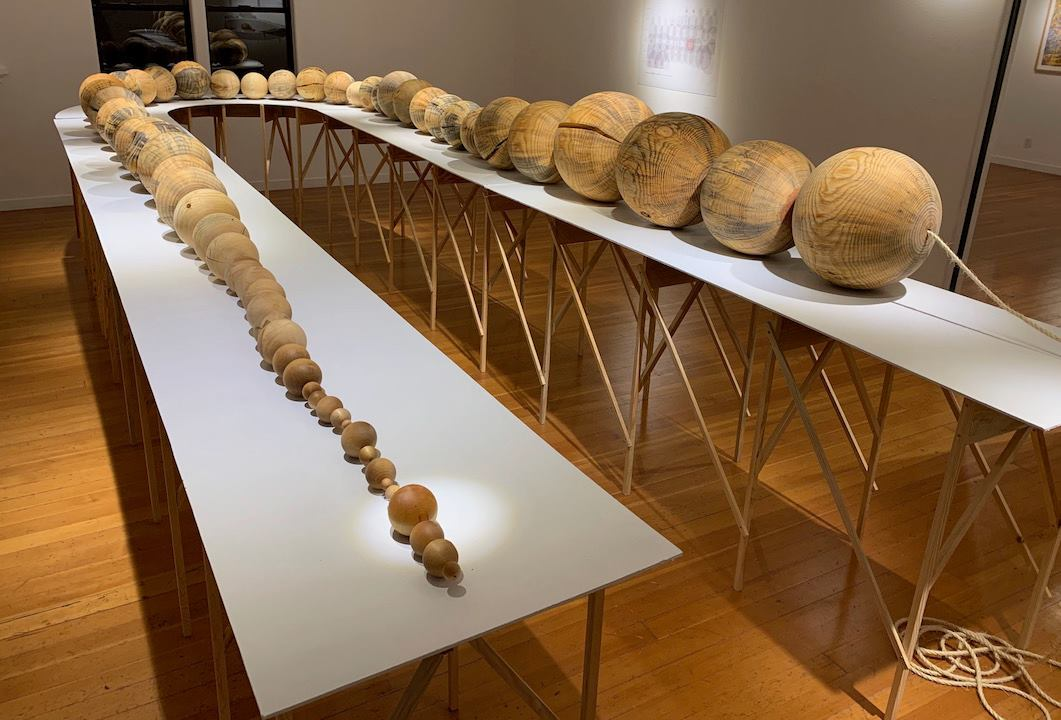}
 \includegraphics[width=0.41\columnwidth]{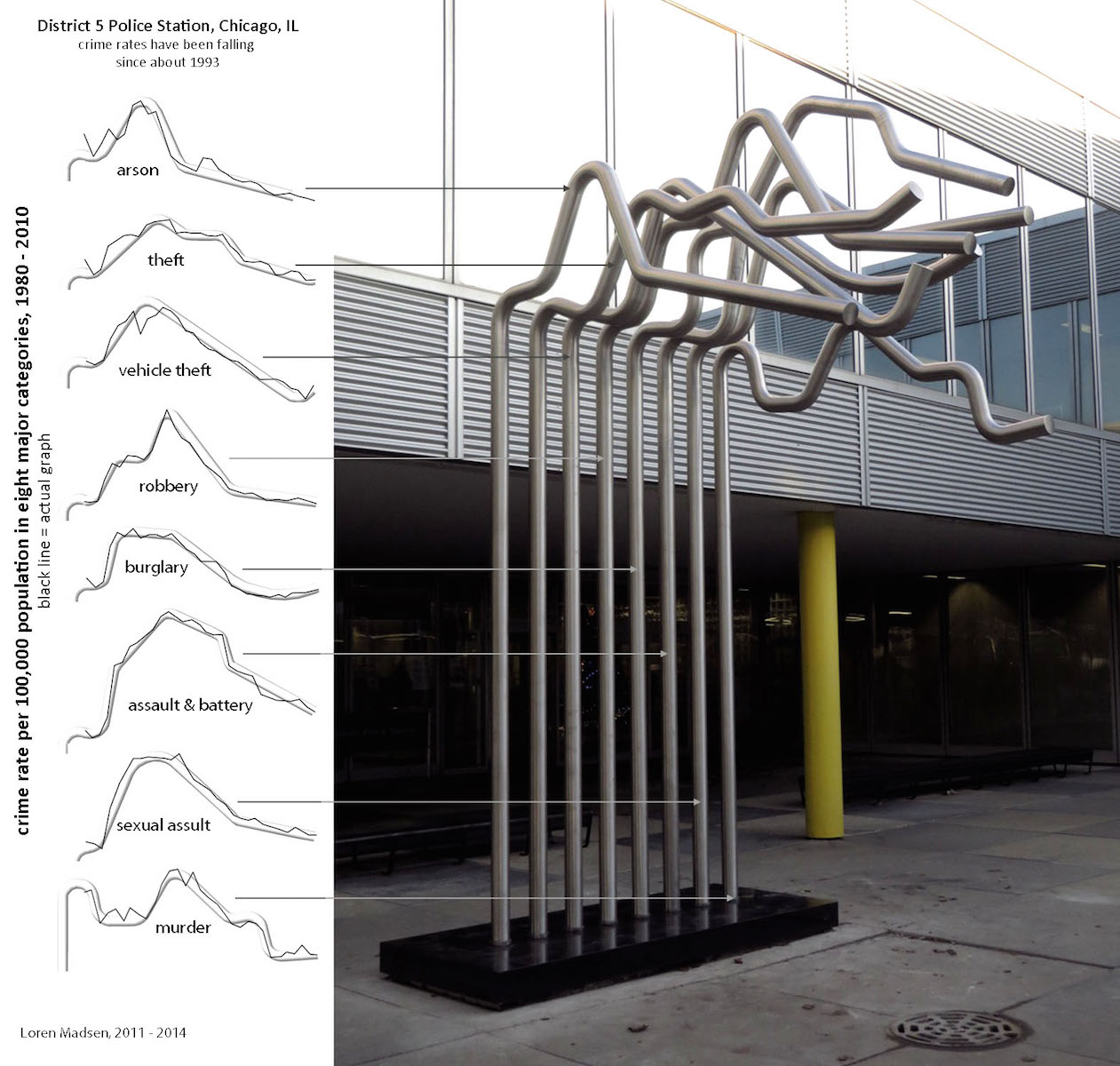}
\caption{Two data sculptures form artist Loren Madsen. \textit{Left:} \textit{Worry Beads} represent 413,000 deaths from terrorism across the world from 1945 (sphere closest in the picture) to 2017 (last sphere on the right). The diameter of each sphere is proportional to the number of people killed that year. \textit{Right:} \textit{District 5}, an outdoor sculpture where steel tubes show falling crime rates across eight crime categories over 30 years. Images from \imgsource{www.lorenmadsen.com}{http://www.lorenmadsen.com/}. \permissions}
 \label{fig:madsen}
\end{figure}

\begin{figure}[tb]
 \centering
 \includegraphics[width=0.49\columnwidth]{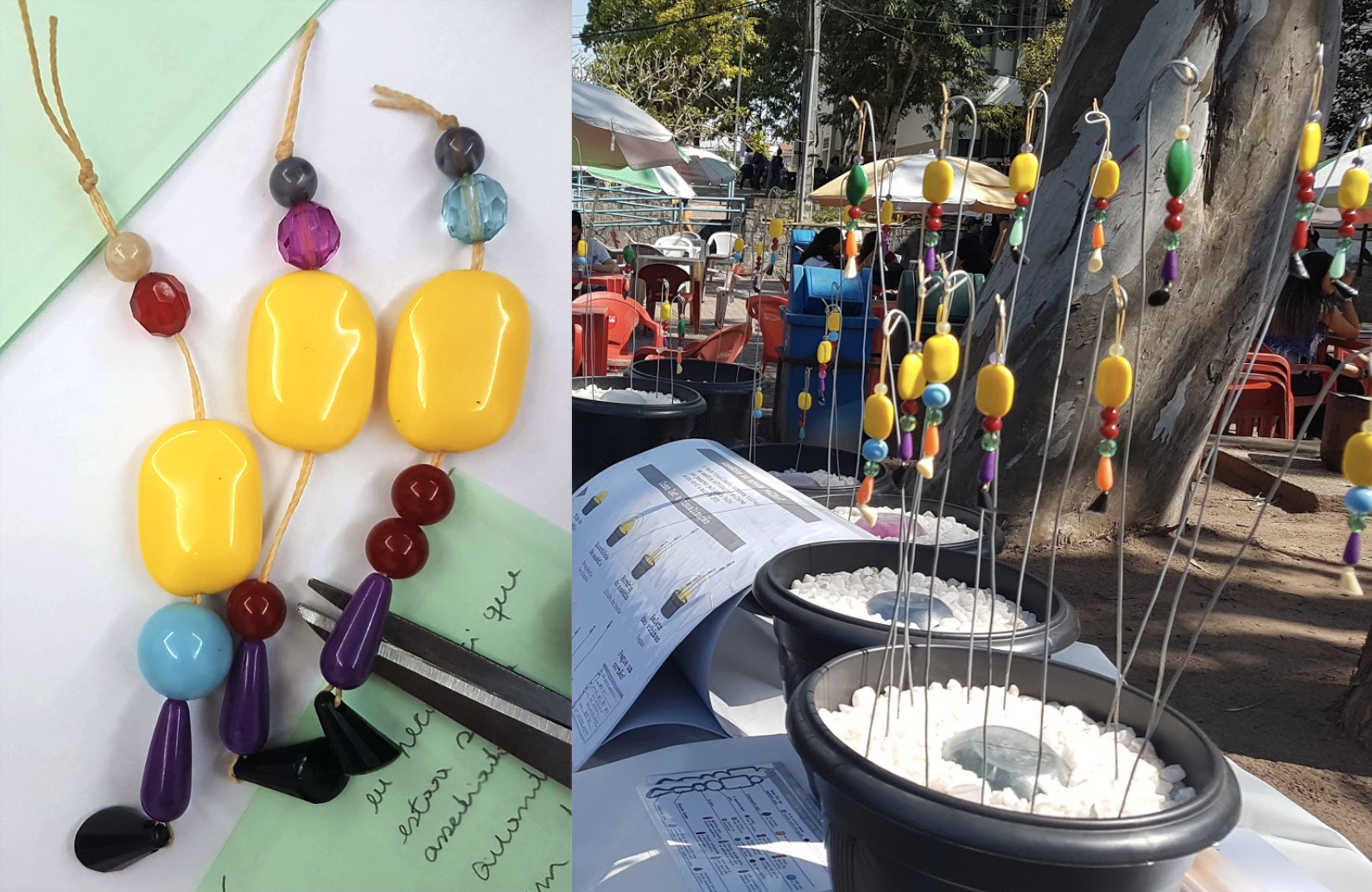}
 \includegraphics[width=0.48\columnwidth]{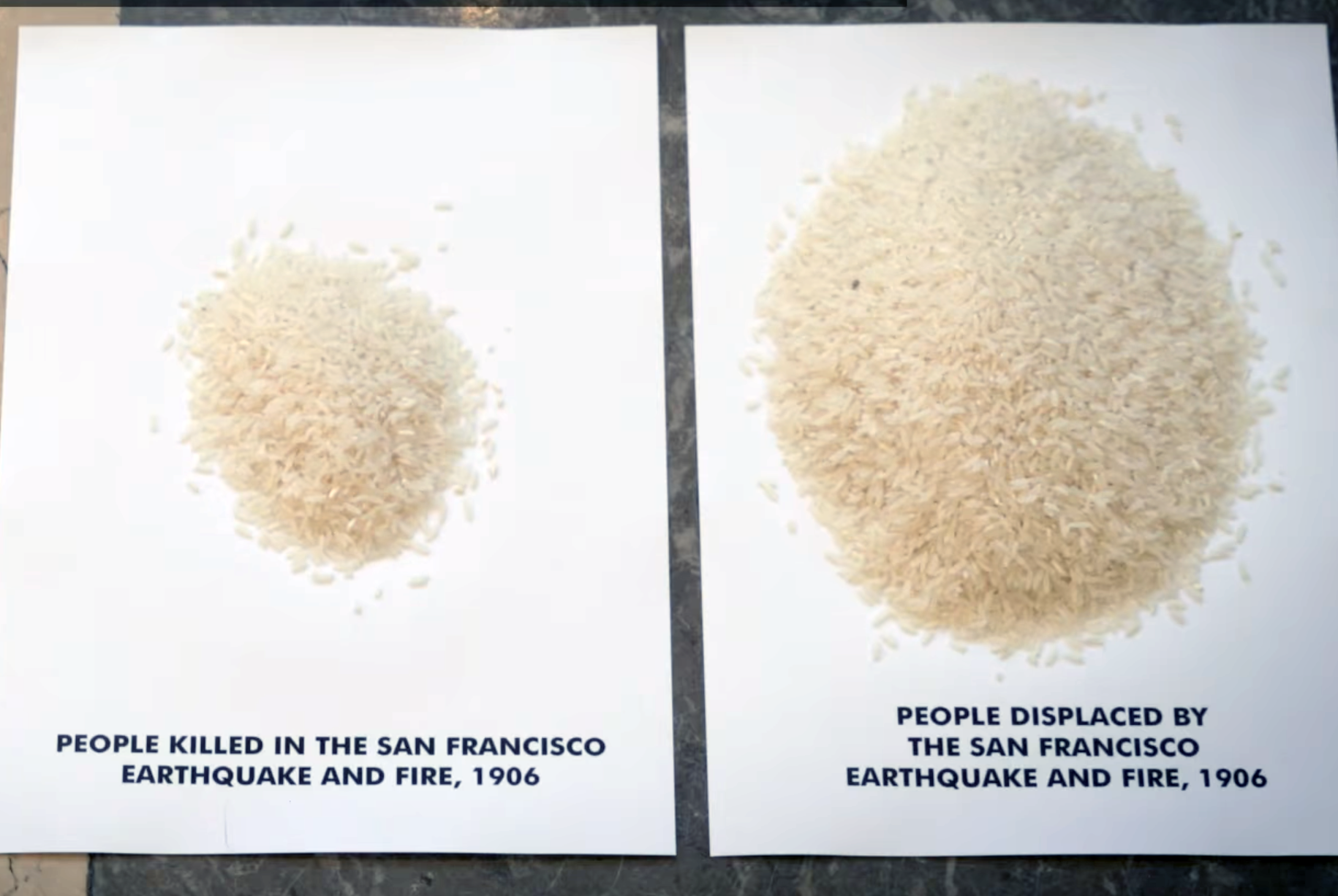}
\caption{\textit{Left:} Physical representations of 28 cases of sexual harassment \textit{Right:} Visualizations of victims of the 1906 earthquake, where each grain of rice represents one victim. Images from \imgsource{luizaugustomm.github.io}{https://luizaugustomm.github.io/pages/harassment-plants.html} and \imgsource{youtube}{https://www.youtube.com/watch?v=zvv110w5lww}. \permissions}
 \label{fig:people-dataphys}
 \vspace{-2mm}
\end{figure}

Although all such physical models focus on conveying qualitative content and visceral experiences by representing real scenes, physical objects can also be used to convey quantitative information. Those displayed in public spaces or museums are frequently called data sculptures \cite{dragicevic2020data}. The artist Loren Madsen, for example, has been creating data sculptures to convey data about social and humanitarian issues since the mid-1990s (\autoref{fig:madsen}) \cite{madsen2015interview}. Compared to the sculptures and dioramas discussed previously, those are much more factual and quantitative. Again there are many ways to bridge the gap between qualitative information and visceral experiences on one side, and quantitative information on the other side. One example of a step in that direction is the data sculpture by Morais et al. \cite{morais2022exploring} (\autoref{fig:people-dataphys}-left). The sculpture, which was exhibited in a public lakeside in Brazil, consists of 28 objects evoking plants, each representing a woman who reported having experienced sexual harassment in the same lakeside. The beads on each plant convey the type of harassment the woman was victim of, the time of the day, the perceived age of the perpetrator(s), and the reaction of the victim. This design is close to the kind of design used in many web-based anthropographics (see Introduction and \autoref{fig:examples-anthropographics}): it shows data, but each person is a separate, recognizable entity. As an example of less abstract data encoding, we could imagine a physical version of the immersive humanitarian visualization by Ivanov et al. \cite{ivanov2018exploration}, where age group and sex are conveyed by the appearance of the sculptures (\autoref{fig:ivanov}). Alternatively, collections of physical objects can be used to convey large numbers of victims, which is a common data sculpture theme (\autoref{fig:people-dataphys}-right)\footnote{Other examples include art installations and memorials where victims are represented by nails (\imgsource{dataphys.org}{http://dataphys.org/list/covid-deaths-as-nails/}), artificial flowers (\imgsource{dataphys.org}{http://dataphys.org/list/tower-poppies-888246-ceramic-poppies-to-commemorate-fallen-soldiers-in-ww1/}), toe tags (\imgsource{dataphys.org}{http://dataphys.org/list/hostile-terrain-94/}), shoes  (\imgsource{cnn.com}{https://edition.cnn.com/2018/03/13/politics/shoe-memorial-congress-gun-violence/index.html}), tattoo dots (\imgsource{dataphys.org}{http://dataphys.org/list/and-counting-tattoo-of-war-casualties/}), and actual people (\imgsource{dataphys.org}{http://dataphys.org/list/mazamet-ville-morte/}, \imgsource{classicfm.com}{https://www.classicfm.com/discover-music/san-francisco-gay-mens-chorus-aids-epidemic/}).}. The design space of humanitarian data sculptures is vast but like representational sculptures, most of them are static. Furthermore, because data sculptures are typically difficult to replicate, they are currently largely confined to public spaces and museums.

\subsection{Ambient Displays}

\begin{figure}[tb]
 \centering
 \includegraphics[width=0.49\columnwidth]{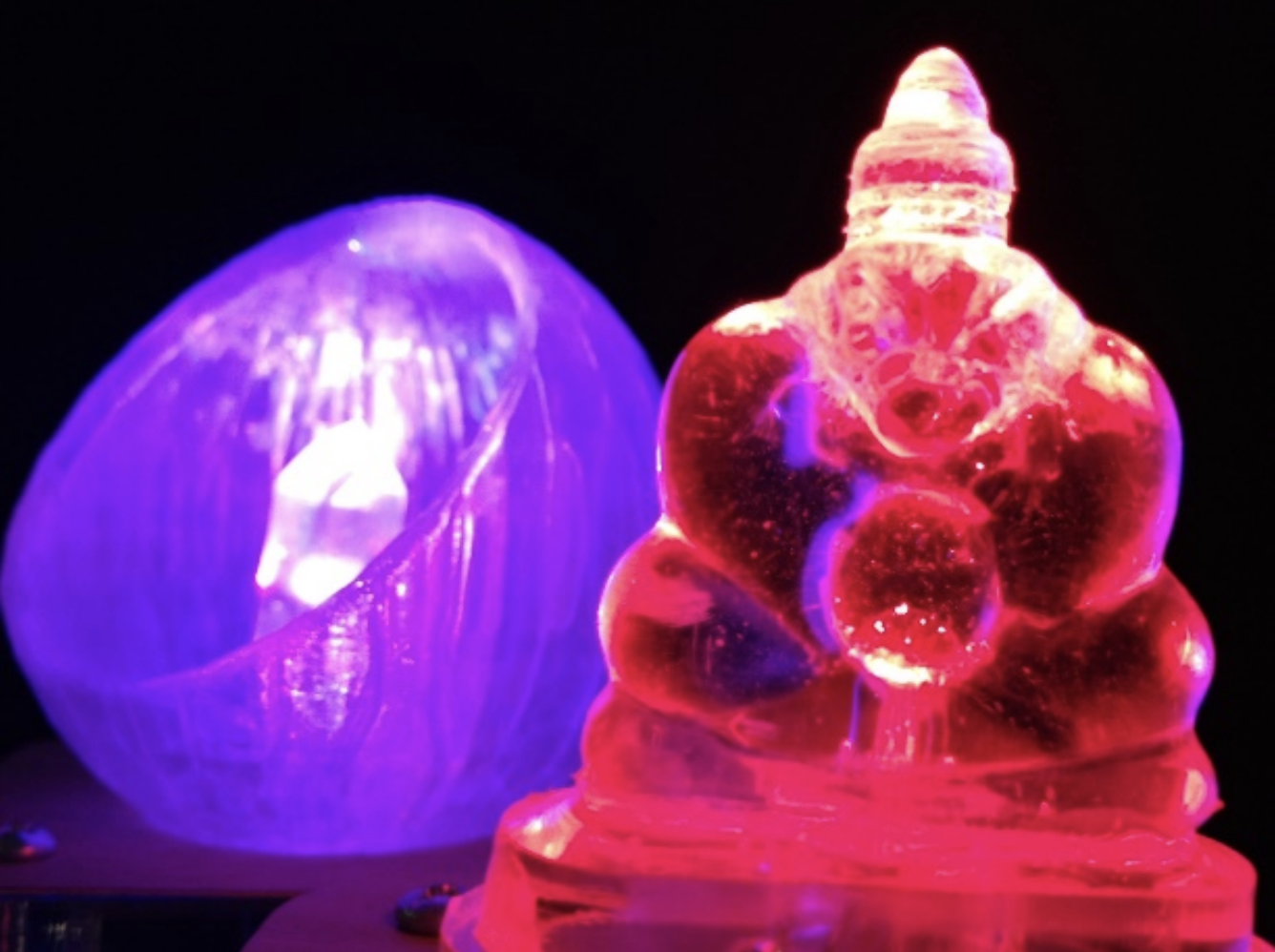}
 \includegraphics[width=0.48\columnwidth]{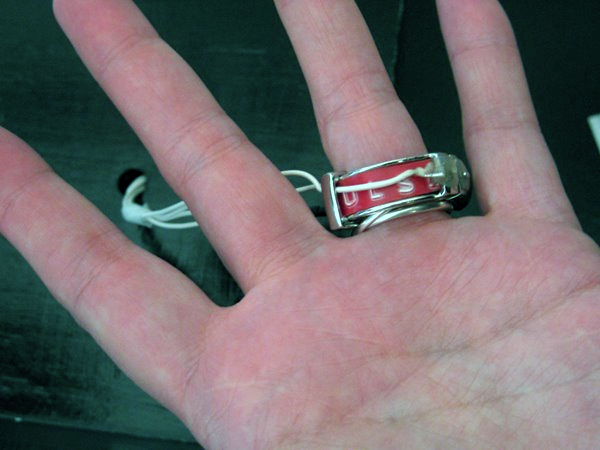}
\caption{Ambient displays for remote intimacy. \textit{Left image}: \textit{BioCrystal} visualizes the affective state of a remote intimate partner, captured through a wearable biofeedback device \cite{roseway2015biocrystal}. The light changes its color from white, to red, green, dark blue, or light blue, depending on the inferred degree of arousal and valence. \textit{Right image}: a ring that measures the pulse of the person wearing it, and transmits it on the partner’s ring as vibration patterns \cite{werner2008united} Image from \imgsource{uxrelatedness.blogspot.com}{http://uxrelatedness.blogspot.com/2013/09/united-pulse.html}. \permissions}
\vspace{-2mm}
 \label{fig:ambient}
\end{figure}

Ambient displays offer another rich space to explore in the area of humanitarian visualization through physically-augmented reality. Wisneski et al.~\cite{wisneski1998ambient} defined them as displays that \myinlinequote{present information within a space through subtle changes in light, sound, and movement, which can be processed in the background of awareness}. To the authors, while personal computing tends to isolate people, ambient displays provide the opportunity to connect them. They instrumented their lab to capture different types of human and animal activity, and transmitted them in real-time to another room through subtle sounds and moving visual patterns on walls and ceilings. They suggested that common house appliances and fixtures can be used for ambient display, and that such displays can be used to monitor loved ones or activities of large groups of people (e.g., stock values, network traffic). Several ideas have been further explored in research, among which the use of ambient displays to support remote intimacy (\autoref{fig:ambient}). Although many designs so far focus on connecting significant others, similar ideas could be used to convey the suffering of distant or anonymous people. For instance, ambient displays could be used for monitoring the severity of an ongoing humanitarian crisis -- such as the number of people currently hospitalized during a severe pandemic, or the likelihood that a war is about to start between two countries, as assessed by prediction markets \cite{luckner2008prediction}. By providing a subtle and continuous impression of an ongoing crisis, ambient displays could serve as a useful complement to the news media. They could reduce the frequency with which people need to poll news reports, which tend to be repetitive and anxiety-inducing, and only occasionally informative.

Although the ambient displays we have seen are minimalist, they can also convey richer data through more complex visualizations. Researchers have experimented with ambient data visualizations, although often with a stronger emphasis on aesthetics than on function \cite{skog2003between, pousman2006taxonomy}. Ambient data visualizations have been suggested for promoting behavior change \cite{moere2007towards}, but mostly for encouraging pro-environmental and healthy habits. The space of ambient visualizations for promoting humanitarian awareness and prosocial behavior remains to this date largely unexplored. In addition, the use of ambient sound to convey data (i.e., data sonification) is interesting to explore as an alternative approach. Among possible sources of inspiration is a project where musicians collaborated with medical doctors to find out how to replace anxiety-inducing hospital alarms by ambient soundscapes that are not only more pleasant, but convey richer and more subtle information about patient status \cite{mars2019sound}.

\section{Promoting Effective Altruism}
\label{sec:effective}
\vspace{1mm}
\fbox{
    \parbox{0.96\linewidth}{\small
2024 update: in the light of revelations about serious problems in some branches of the effective altruism movement, I no longer support the movement as a whole 
(more info at \href{https://dragice.fr/ea.html}{https://dragice.fr/ea.html}).
}}
\vspace{1mm}

Here I argue that to be really effective, future immersive humanitarian visualizations will need to be based on the principle of effective altruism. But before I explain effective altruism, I will first motivate its importance by discussing the limits of storytelling.

\subsection{Effective Stories vs. Effective Decisions}
\label{sec:stories-decisions}

\begin{figure}[tb]
 \centering
 \includegraphics[width=0.9\columnwidth]{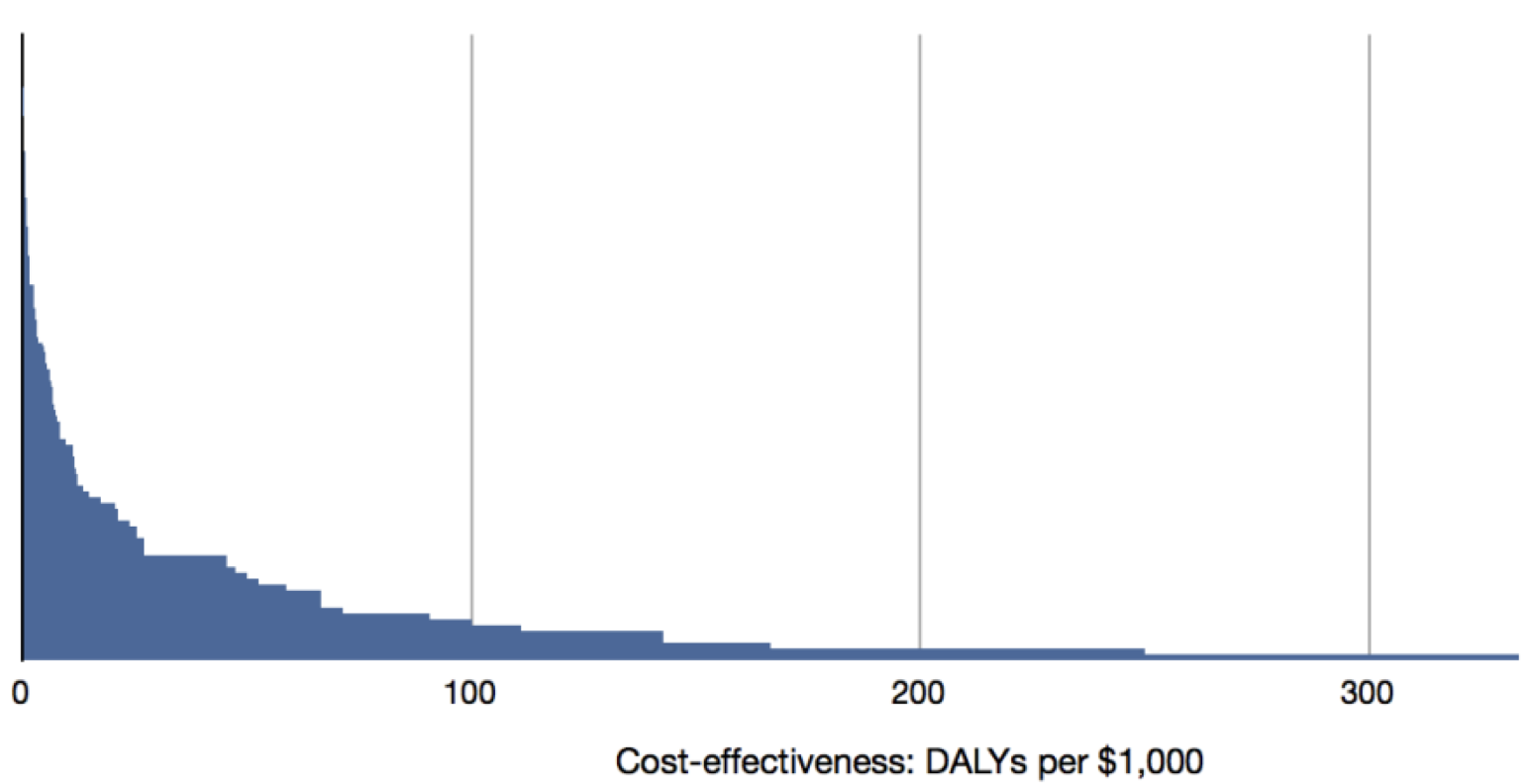}
\caption{Distribution of the cost-effectiveness of 100+ health interventions, expressed in DALYs per \$1,000. DALYs means disability-adjusted life years, and it is a unit that combines the number of years of life saved and the number of years of disease prevented by the intervention. Image from \cite{ord2013moral}. \permission}
 \label{fig:cost}
 \vspace{-2mm}
\end{figure}

Among the speculative immersive humanitarian visualization systems I discussed, several involve telling personal stories. According to Neil Halloran, the designer of the celebrated data-driven documentary ``the fallen of WWII'' (\autoref{fig:examples-anthropographics}-right), personal stories are very powerful and many data journalists and designers recommend using them to maximize impact \cite[starts at 23 min]{halloran2017emotional}. However, in his keynote presentation, Halloran admits that he hates this advice, and that he prefers to elicit emotions without relying too much on human stories. A major limitation of stories is that \myinlinequote{you can tell a story about a crisis of any size, and tell a compelling story} \cite[at 25 min]{halloran2017emotional}. Because of this, individual stories and perhaps storytelling in general are insufficient to help people think rationally about human suffering and how to best alleviate it. To illustrate, consider for example that with \$40,000, we can train and provide a dog for a blind person in the U.S, or cure more than 2,000 people of blindness by paying for surgeries to reverse the effects of trachoma in Africa \cite{ord2013moral}. While most people would agree that the latter is by far the best choice, it is possible to tell a compelling human story in both cases, so stories alone provide poor guidance. To appreciate how general this problem is, consider the distribution of the cost-effectiveness of 100+ health interventions shown in \autoref{fig:cost} \cite{ord2013moral}. Effectiveness spreads over more than 4 orders of magnitude, ranging from 0.02 to 300 DALYs per \$1,000, with a median of 5, which is an extremely skewed distribution. An important consequence is that moving money from the many ineffective interventions to the most effective ones is likely to be helping people considerably more than donating even a lot of money to a random intervention.


This way of thinking about charity as an endeavor that strives to maximize positive impact (as opposed to, e.g., make donors feel good) has been called \textit{effective altruism} \cite{macaskill2020can,singer2013why}, and it is a movement that is becoming more and more important. In my view, it would be highly beneficial if many future immersive humanitarian visualizations were informed by effective altruism.

\subsection{Supporting Informed Comparisons}

Promoting effective altruism would mean among other things more emphasis on visualizations supporting comparisons rather focusing on single issues or single charities. Visualizations supporting comparisons could help make more egalitarian funds allocations -- for example, if a deadly disease affects 400 people in a country and 40,000 people in another and curing a person costs the same, a decision maker should give the latter country about 100 times more funding than the former. To help individual people choose charities and causes to support, data is available on effective altruism websites, although few visualizations are currently available. For example, GiveWell maintains a list of top effective charities, primarily based on the cost of life saved (which is about \$4,500 for top charities like the Against Malaria Foundation) \cite{givewell2022top}. 

Because people are most often exposed to humanitarian issues on a case-by-case basis and rarely find themselves in situations involving comparisons, it would be ideal if visualizations could promote effective altruism by design, without requiring comparisons. Unfortunately, this seems extremely difficult to achieve. Such a visualization would require that if someone decides one day to donate a certain sum to alleviate a tragedy, if the next day they encounter a tragedy $N$ times as bad, they should donate $N$ times more. Not only this is a financial impossibility, this would also be a psychological impossibility if donations were motivated by empathy or compassion -- if a person is capable of feeling sorry or concerned about one person's death or suffering, it would be psychologically impossible for them to feel 10,000 times more sorry or concerned about the same plight affecting 10,000 people \cite{fetherstonhaugh1997insensitivity}. Therefore, it seems that the most straightforward way for visualizations to promote effective altruism is by being very good at supporting comparisons and the fair allocation of resources across options. However, this is not without difficulties either, for reasons we will now see.

\subsection{The Challenges of Conveying Outcomes}
\label{sec:outcomes}

Although some of the data from effective altruist organizations can be visualized, important visualization design challenges arise when outcomes other than death counts need to be visualized and compared. For example, about 600 mosquito nets prevent the death of a child, but they also prevent 500 to 1,000 cases of malaria \cite{against2022why}. This is an enormous benefit in and of itself, as malaria is a crippling disease with flu-like symptoms that can periodically return, can be highly disruptive for the life of households, and can leave children disabled \cite{ricci2012social}. Similarly, GiveWell lists a charity that saves lives by giving vitamin A supplements to children, but even when it is not fatal, vitamin A deficiency causes a range of terrible problems such as repetitive infections and blindness \cite{world2022vitamin}.
Another effective altruism website lists a charity that can use about \$20,000 to prevent a year of homelessness in the US or UK \cite{clare2020homelessness}, and another one that can use \$200--\$300 to prevent the equivalent of one year of severe major depressive disorder for a woman in Uganda \cite{halstead2019mental}. It is very hard to imagine how to visualize those widely different types of outcomes in a way that supports informed, effective-altruist decisions.

\begin{figure}[tb]
 \centering
 \includegraphics[width=0.45\columnwidth]{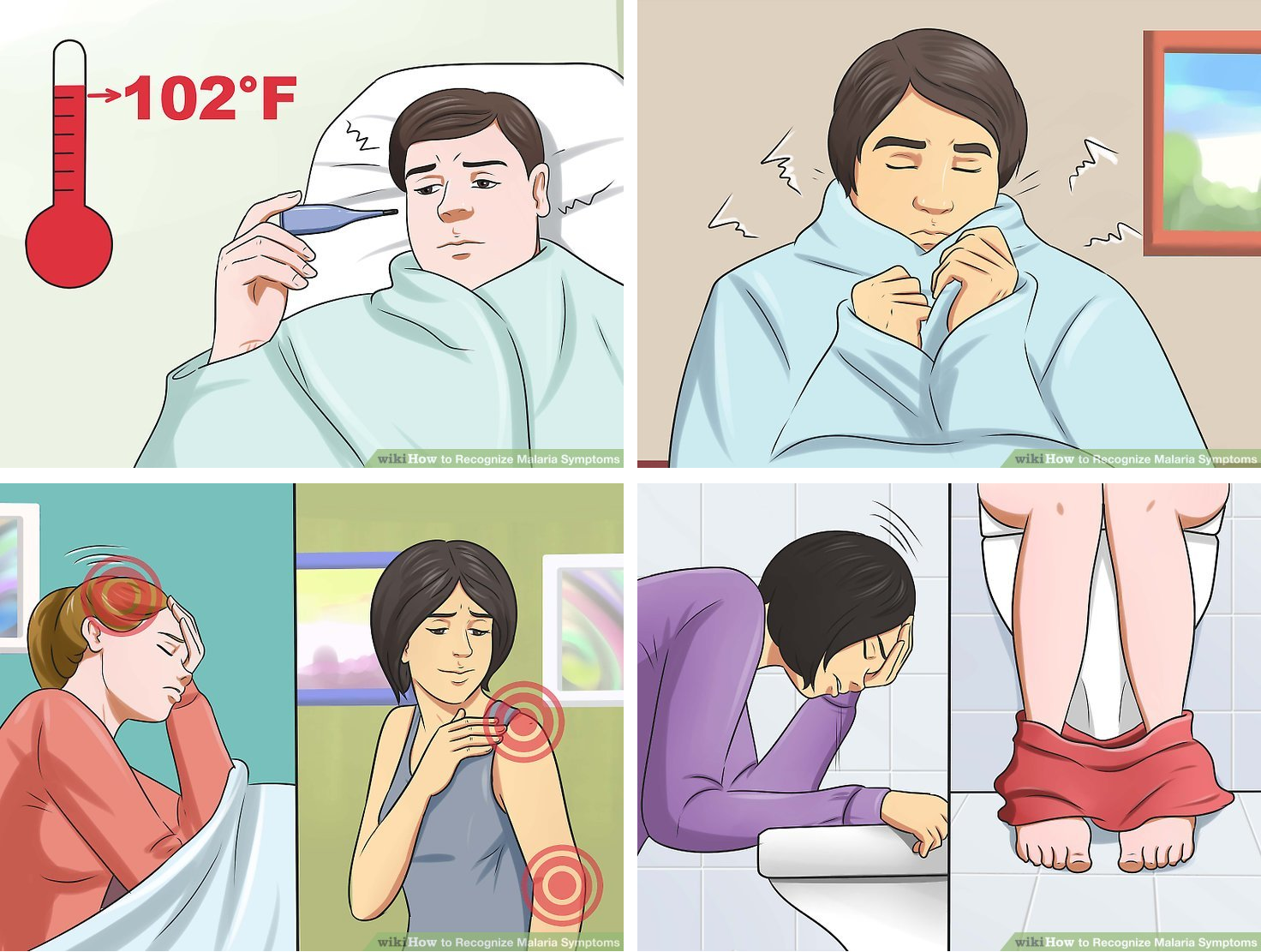}
 \includegraphics[width=0.53\columnwidth]{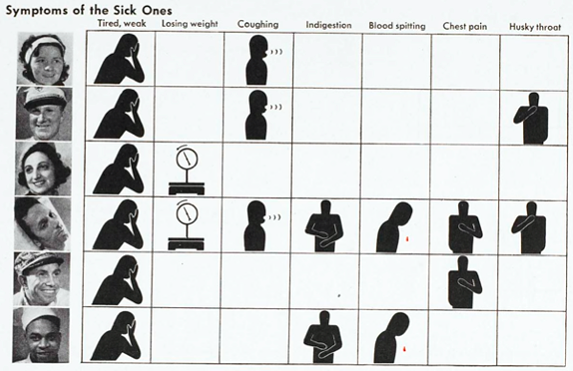}
\caption{Visual illustrations of disease symptoms. \textit{Left:} malaria symptoms; \textit{Right:} different possible combinations of tuberculosis symptoms. Image sources:  \imgsource{www.wikihow.com}{https://www.wikihow.com/Recognize-Malaria-Symptoms} and \imgsource{designobserver.com}{https://designobserver.com/feature/birthing-dataviz/40246/} (original source International Foundation for Visual Education, Tuberculosis, 1939). \permissions}
 \label{fig:symptoms}
\end{figure}

Immersive humanitarian visualizations offer the opportunity to address these hard issues due to their ability to combine the communication of quantitative, factual data with the communication of qualitative information and visceral experiences. Ideally, an important donation or funds allocation decision would be based both on quantitative facts (e.g., the number of people affected, the cost of interventions) and a deep understanding of people's subjective experiences with and without the interventions, especially concerning the degree of physical and psychological suffering involved. However, it is hard for a person who has never contracted malaria or never had a vitamin A deficiency to have a reliable intuition of what those experiences entail. This is where immersive storytelling could be useful. Similar to the hypothetical war map example discussed in \autoref{sec:unfolding-tragedies}, one could imagine interactive visualizations showing quantitative facts and simulated outcomes, and where viewers can zoom on individual people to immerse themselves into their real or hypothetical subjective experience. It is challenging to reconcile the world of numbers with the world of subjective experience, but not entirely impossible -- for example, if an effective altruist judges that having disease A is twice as bad as having disease B, they could conclude that preventing 10,000 cases of disease A is equally desirable as preventing 20,000 cases of disease B.\footnote{One key limitation is our inability to fully understand suffering without directly experiencing it; The intensity of physical and psychological pain can vary by orders of magnitude across different experiences, with no clear upper limit \cite[Chap.~4]{vinding2020suffering}. Thus it is unclear if  only witnessing or experiencing a tiny fraction of the pain is sufficient to be able to compare pain intensities.} However, getting immersed in stories takes time, and simpler formats may be needed to communicate outcomes when large datasets need to be explored, or when quick decisions are needed. A possible approach could be to first get the decision maker immersed in stories to get a visceral understanding of the ranges of experiences involved, and, at anytime after that, let them explore data using visualizations combined with visual summaries referring to those subjective experiences (a \textit{stories-then-data} approach similar to Hans Rosling's magic washing machine, see \autoref{sec:positive}). For the visual summaries, perhaps there is inspiration to be found in the various diagrams and illustrations used to convey pain scales and disease symptoms (\autoref{fig:symptoms}). 

\begin{figure}[tb]
 \centering
 \includegraphics[width=1.0\columnwidth]{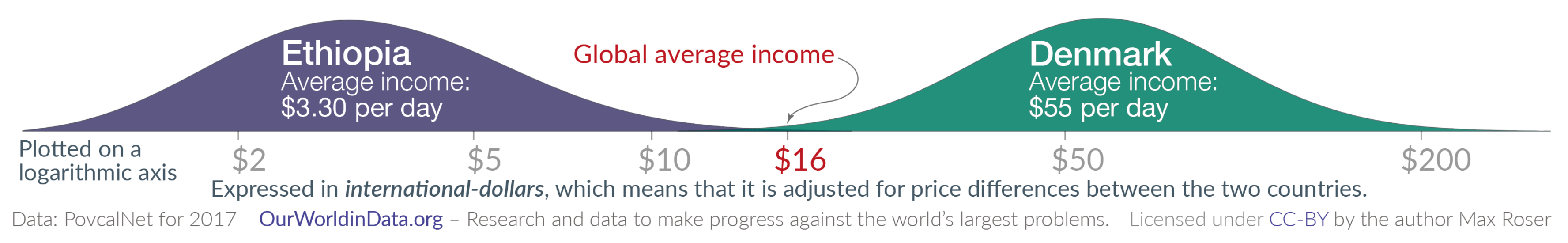}
\caption{Distribution of incomes in Ethiopia and Denmark, adjusted for price differences. A person born in Denmark is almost certain to have an income way above the global average, while a person born in Ethiopia is virtually guaranteed to have an income way below. Image source \imgsource{ourworldindata.org}{https://ourworldindata.org/global-economic-inequality-introduction}. Image CC BY 4.0.}
 \label{fig:inequality}
 \vspace{-2mm}
\end{figure}

\subsection{Connecting Data with our Everyday Lives}

Besides helping allocate resources to different humanitarian actions, comparison-focused immersive humanitarian visualizations could help viewers compare their own condition with the condition of less advantaged people. Several visualizations and infographics on global inequalities were already designed to show that \myinlinequote{where a person finds themselves in the extremely unequal global income distribution is mostly determined by where they are} (\autoref{fig:inequality}). By adding a qualitative and experiential component to this type of factual information (see for example the Dollar Street app on \autoref{fig:dollar-street}), immersive humanitarian visualizations could help viewers get a deeper understanding of the importance of focusing on distant rather than local people and of promoting global resource redistribution.

\begin{figure}[tb]
 \centering
 \includegraphics[width=0.95\columnwidth]{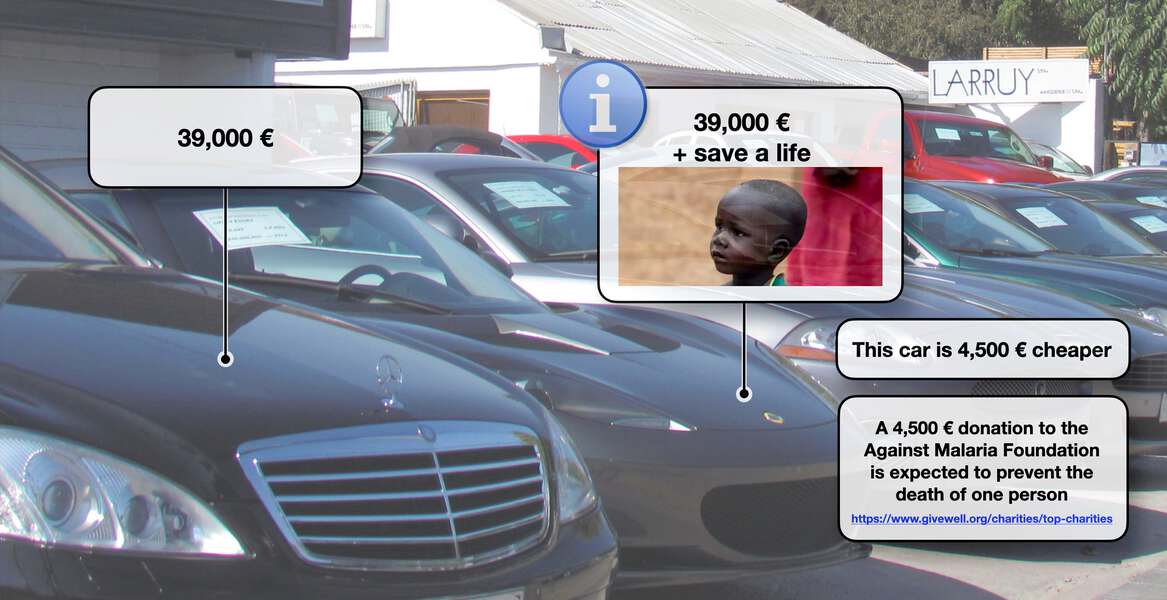}
\caption{Mock-up of a possible future \ar system informing people of ways of redirecting unnecessary personal expenses for humanitarian purposes. Background photo from \imgsource{commons.m.wikimedia.org}{https://commons.m.wikimedia.org/wiki/File:Used_car_dealership_in_Santiago,_Chile.jpg}. Both background photo and this image under CC BY-SA 2.0.}
 \label{fig:cars}
 \vspace{-2mm}
\end{figure}

Finally, comparison-focused immersive humanitarian visualizations could help people better understand the potential impact of their everyday choices on the welfare of other individuals, and raise their awareness about alternative choices. Augmented-reality and ambient displays are especially interesting to explore for this purpose. For example, an \ar system could inform us that by spending one less day of vacation in a family resort or by giving up on an expensive pair of sunglasses, we can prevent a full year of severe major depression. Or it could inform us that by saving 4,500€ on a car, we could be driving a slightly less powerful car, but with the satisfaction that we have probably saved a life (\autoref{fig:cars}). This mock-up is only a starting point for reflection, and it could be interesting to think about how such systems could be complemented with rich data visualizations and more elaborate outcome representations (e.g., to convey non-death related outcomes as discussed in \autoref{sec:outcomes}).

\section{Concluding Remarks}

This overview of the possibilities offered by immersive humanitarian visualizations was more a stroll through possible directions of research than a walk through a comprehensive design space. A systematic mapping of this rich design space is one of the great research tasks that will need to be accomplished and that this preliminary exploration will hopefully motivate. In what follows I identify other large gaps in this essay and suggest ways to fill them.

\subsection{Related Research Areas}

There is a vast body of related work that can inform research in this area, and of which I have barely scratched the surface here. I have already mentioned related areas in visualization and human-computer interaction, such as  research on anthropographics \cite{boy2017showing, morais2020showing, morais2021can}, narrative visualization and storytelling \cite{segel2010narrative, kosara2013storytelling, riche2018data}, visualization ethics \cite{correll2019ethical}, immersive analytics \cite{marriott2018immersive}, situated visualizations \cite{willett2016embedded, bressa2021s}\footnote{Note that in the \ar relocation scenarios I discussed, the \ar representations are not situated in the sense of Willett et al., \cite{willett2016embedded} because they are far away from their physical referent. However, they are situated in the broader sense put forward by Bressa et al. \cite{bressa2021s}.}, and ambient displays and visualizations \cite{wisneski1998ambient, skog2003between, pousman2006taxonomy}. Other relevant research areas include persuasive technologies \cite{hamari2014persuasive}, visualizations for healthcare \cite{shneiderman2013improving} and visualizations of public health data (e.g., \cite{zhang2021mapping}). There is also highly relevant literature outside visualization and human-computer interaction, especially in psychology and behavioral economics. I have already mentioned studies on empathy and perspective-taking in \vr \cite{herrera2018building, martingano2021virtual}. Also deeply relevant are studies on altruism and charitable giving \cite{bloom2017empathy, noetel2020we}, a few of which have been conducted in \vr environments (e.g., \cite{kandaurova2019effects, calvert2019design}). Work on moral philosophy \cite{singer1972morality, singer2011expanding, vinding2020suffering} and on narrative journalism \cite{sillesen2015journalism} is also relevant. Any survey that dives deeply in one or several of these areas and connects them to immersive humanitarian visualization would make an important research contribution.

\subsection{Ethical Issues}
\label{sec:ethics}

Ethical issues will be important to consider when designing and deploying immersive humanitarian visualizations. Foreseeable use cases involve at least two parties: victims or people needing help about whom data or information is communicated, and viewers who may be in position to help. It is important that both parties are respected and that their informed consent is obtained. On the viewer's side, being exposed to human suffering and death can be distressing and carries psychological risks of which the viewer needs to be made aware of. Even more importantly, people about whom information is communicated (or their families if they are deceased) must spontaneously volunteer this information or at the very least agree for it to be communicated. But even then, their information can be misused, e.g., for voyeurism or sensationalism. Voyeurism (watching others suffer out of curiosity or for entertainment) seems unfortunately hard to prevent. However, sensationalism (reporting to shock rather than to inform) works by selecting extreme aspects of reality \cite{meikle2013poverty}, which can be avoided in a data-driven approach focusing on conveying the full range of human experiences on a given issue and how common they are.

Some of the technologies that can enable immersive humanitarian visualizations carry risks in and of themselves. For example, with future lightweight wearable \ar displays, it is important that people are able to distinguish reality from computer renditions and can easily disable their devices anytime. Furthermore, immersive humanitarian visualization is vulnerable to nefarious uses and dark design patterns \cite{gray2018dark, correll2019ethical} that can go from excessive nudging by charity organizations to the hijacking of the technology by totalitarian regimes for large-scale propaganda and misinformation. For example, both quantitative and qualitative information could be manipulated to change people's attitude about an unfolding war. Arguably propaganda and misinformation are already possible with current media and technologies, but if immersive humanitarian visualizations are really more effective at conveying true information, they will be also more effective at conveying false information.

\subsection{Issues of Social Distance}
\label{sec:social-distance}

I discussed how immersive technologies can reduce the perceived spatial distance between viewers and victims, but future work will need to consider other types of psychological distances such as temporal distance and social distance \cite{liberman2007psychological}.\footnote{Similarly, Phelps \cite{phelps2021killing} defines the total distance between a killer and their remote victim as the sum of physical distance, cultural distance, moral distance, social distance, mechanical distance, and empathetic distance.} Emotional responses tend to decrease with psychological distance, with many studies specifically suggesting that empathy fades with social distance \cite{liberman2007psychological}. Social distance includes, e.g., whether the victim is a friend or a stranger, and whether they share the same culture. This poses a challenging problem to humanitarian visualization, as communicating more information about victims could emphasize the extent to which they may differ from the viewer in terms of their culture, political leaning, religion, or race. The response from charity organizations has sometimes been to withhold information: for example, in their experimental donor-recipient pairing program, GiveDirectly mentioned that the pairing is done by them because \myinlinequote{when donors choose, the evidence is that people choose based on things like the color of someone's skin or how attractive they are, and that doesn't make for particularly good social policy or operations} \cite{faye2021cash}. Similarly, the Tomb of the Unknown Soldier conveys as little information as possible so that anyone could relate to the solider and entertain the possibility that they are a lost relative \cite{mars2019known}. However, it has also been suggested that immersive technologies can help overcome individual differences (e.g., \cite{peck2013putting}). Thus, it is possible that when revealing information about victims (e.g., by showing photo portraits), empathy and concern may in some cases decrease, but as we share more and more information and allow viewers to immerse themselves into the lives of the victims and connect with their humanity, empathy and concern increase again and rise above the original level where little or no information is shared.

\subsection{Non-Human Animals}

Finally, humanitarian visualizations are about promoting human welfare, but our society is increasingly concerned about the welfare of non-human animals. This expansion of society's moral circle \cite{singer2011expanding} has started with animals whose suffering is caused by humans (mostly, farmed animals) and will likely extend to all animals \cite{tomasik2018trc,  soryl2021case}. Therefore, it is useful to move beyond humanitarian visualizations and consider the study of \textit{sentiocentrist visualizations}, which are designed to promote the welfare of all sentient individuals.\footnote{\myinlinequote{Sentiocentrism, sentio-centrism, or sentientism is an ethical view that places sentient individuals (i.e., basically conscious beings) at the center of moral concern.} \url{https://en.wikipedia.org/wiki/Sentiocentrism}}

While some of my discussions and proposed designs for immersive humanitarian visualizations likely generalize to non-human animals, most non-human animals are quite different from us and their ability to communicate with us is very limited, so it is harder to empathize with them. There has been some progress in using immersive display technologies to help bridge this empathy gap, for example by using \vr to emulate contact with wild animals who are otherwise difficult to access \cite{pimentel2021peril} or to elicit the experience of inhabiting the body of an animal \cite{ahn2016experiencing}, or by producing 360-degree video documentaries shot from the perspective of farmed animals \cite{ianimal}. On the data visualization side, one possible source of inspiration for immersive sentiocentrist visualizations is Sauvé et al.'s \textit{Econundrum} project \cite{sauve2020econundrum}, which is an ambient data sculpture showing the impact of dietary choices made by a small community. Although it visualizes carbon emissions, this type of approach could be re-purposed to show the impact of dietary choices on animal suffering. The difficulty of assessing animal suffering across species poses a great challenge, but it is possible to make educated guesses \cite{tomasik2019brain} and reason with incomplete information \cite{tomasik2018how}.

Just like for immersive humanitarian visualizations, it is important that future sentiocentrist visualizations are informed by effective altruism. Such visualizations could use data to encourage clear thinking about animal welfare, e.g., by revealing possible cognitive biases associated with the topic\footnote{To take an example, veganism and vegetarianism tend to encourage a purist, often quasi-religious mind frame that has no direct benefit to animals -- for example, it is twice as good for animals if four people decide to cut meat consumption by half than if a single person decides to cut meat entirely.} or by exposing tensions between animal welfare and other causes like environmentalism \cite{faria2019s,tomasik2018trc,soryl2021case}.

\subsection{Conclusion}

This paper introduced immersive humanitarian visualization as a promising research area in information visualization. It is far from being a comprehensive overview, and many of its ideas are highly speculative. But I hope it will encourage, motivate, and inspire future research on this important topic.

\section*{Acknowledgments}

This paper owes to Luiz Morais, with whom I started working on anthropographics back in 2018. The discussions I had with him and our collaborators Yvonne Jansen and Nazareno Andrade over those years were a great source of inspiration. I was also inspired by recent discussions with Martin Hachet, Arnaud Prouzeau, Ambre Assor and Yvonne in the context of a PhD thesis and a grant proposal which have conceptual overlaps with this paper. Finally, I am very grateful to Yvonne, Wesley Willett, and Theophanis Tsandilas for their excellent comments and suggestions about this paper.

\bibliographystyle{abbrv-doi}

\bibliography{bib-articles,bib-misc}

\begin{thebibliography}{10}

\bibitem{against2022why}
{Against Malaria Foundation}.
\newblock Why nets?
\newblock Web page \url{https://www.againstmalaria.com/WhyNets.aspx}. Last
  visited 2022-03-22., 2022.

\bibitem{ahn2016experiencing}
S.~J. Ahn, J.~Bostick, E.~Ogle, K.~L. Nowak, K.~T. McGillicuddy, and J.~N.
  Bailenson.
\newblock Experiencing nature: Embodying animals in immersive virtual
  environments increases inclusion of nature in self and involvement with
  nature.
\newblock {\em Journal of Computer-Mediated Communication}, 21(6):399--419,
  2016.

\bibitem{ianimal}
{Animal Equality}.
\newblock ianimal.
\newblock Web page \url{https://ianimal360.com/}. Last visited 2022-03-22.,
  2022.

\bibitem{bevan2018history}
C.~Bevan and D.~Green.
\newblock A history of virtual reality nonfiction 2012--2018.
\newblock Web page \url{http://vrdocumentaryencounters.co.uk/vrmediography/}.
  Last visited 2022-03-22., 2018.

\bibitem{bevan2018mediography}
C.~Bevan and D.~Green.
\newblock A mediography of virtual reality non-fiction: Insights and future
  directions.
\newblock In {\em Proceedings of the 2018 ACM International Conference on
  Interactive Experiences for TV and Online Video}, pp. 161--166, 2018.

\bibitem{bloom2017empathy}
P.~Bloom.
\newblock Empathy and its discontents.
\newblock {\em Trends in cognitive sciences}, 21(1):24--31, 2017.

\bibitem{bolt2021effects}
E.~Bolt, J.~T. Ho, M.~Roel~Lesur, A.~Soutschek, P.~N. Tobler, and
  B.~Lenggenhager.
\newblock Effects of a virtual gender swap on social and temporal
  decision-making.
\newblock {\em Scientific Reports}, 11(1):1--15, 2021.

\bibitem{boy2017showing}
J.~Boy, A.~V. Pandey, J.~Emerson, M.~Satterthwaite, O.~Nov, and E.~Bertini.
\newblock Showing people behind data: Does anthropomorphizing visualizations
  elicit more empathy for human rights data?
\newblock In {\em Proc. CHI}, pp. 5462--5474, 2017.

\bibitem{bressa2021s}
N.~Bressa, H.~Korsgaard, A.~Tabard, S.~Houben, and J.~Vermeulen.
\newblock What's the situation with situated visualization? a survey and
  perspectives on situatedness.
\newblock {\em IEEE Transactions on Visualization and Computer Graphics},
  28(1):107--117, 2021.

\bibitem{bryan2010commitment}
G.~Bryan, D.~Karlan, and S.~Nelson.
\newblock Commitment devices.
\newblock {\em Annu. Rev. Econ.}, 2(1):671--698, 2010.

\bibitem{calvert2019design}
J.~Calvert, R.~Abadia, and S.~M. Tauseef.
\newblock Design and testing of a virtual reality enabled experience that
  enhances engagement and simulates empathy for historical events and
  characters.
\newblock In {\em 2019 IEEE Conf. on Virtual Reality and 3D User Interfaces
  (VR)}, pp. 868--869. IEEE, 2019.

\bibitem{caplan2019open}
B.~Caplan and Z.~Weinersmith.
\newblock {\em Open borders: the science and ethics of immigration}.
\newblock First Second, 2019.

\bibitem{card1999readings}
M.~Card.
\newblock {\em Readings in information visualization: using vision to think}.
\newblock Morgan Kaufmann, 1999.

\bibitem{clare2020homelessness}
S.~Clare.
\newblock Homelessness in the {US} and {UK} executive summary.
\newblock Web article
  \url{https://www.founderspledge.com/stories/homelessness-in-the-us-and-uk-executive-summary},
  2020.

\bibitem{caters2016vr}
C.~Clips.
\newblock Vr goggles gender swap experience.
\newblock Video \url{https://www.youtube.com/watch?v=WCprfChibTE}., 2016.

\bibitem{concannon2020brooke}
S.~Concannon, N.~Rajan, P.~Shah, D.~Smith, M.~Ursu, and J.~Hook.
\newblock Brooke leave home: Designing a personalized film to support public
  engagement with open data.
\newblock In {\em Proceedings of the 2020 CHI Conference on Human Factors in
  Computing Systems}, pp. 1--14, 2020.

\bibitem{correll2019ethical}
M.~Correll.
\newblock Ethical dimensions of visualization research.
\newblock In {\em Proceedings of the 2019 CHI Conference on Human Factors in
  Computing Systems}, pp. 1--13, 2019.

\bibitem{dragicevic2020data}
P.~Dragicevic, Y.~Jansen, and A.~Vande~Moere.
\newblock Data physicalization.
\newblock {\em Handbook of Human Computer Interaction}, pp. 1--51, 2020.

\bibitem{elmqvist2013ubiquitous}
N.~Elmqvist and P.~Irani.
\newblock Ubiquitous analytics: Interacting with big data anywhere, anytime.
\newblock {\em Computer}, 46(4):86--89, 2013.

\bibitem{faria2019s}
C.~Faria and E.~Paez.
\newblock It’s splitsville: why animal ethics and environmental ethics are
  incompatible.
\newblock {\em American Behavioral Scientist}, 63(8):1047--1060, 2019.

\bibitem{faye2021cash}
M.~Faye and J.~Galef.
\newblock Is cash the best way to help the poor?
\newblock Rationally Speaking podcast episode \#263
  \url{http://rationallyspeakingpodcast.org/263-is-cash-the-best-way-to-help-the-poor-michael-faye/},
  2021.

\bibitem{fekete2008value}
J.-D. Fekete, J.~J.~v. Wijk, J.~T. Stasko, and C.~North.
\newblock The value of information visualization.
\newblock In {\em Information visualization}, pp. 1--18. Springer, 2008.

\bibitem{fetherstonhaugh1997insensitivity}
D.~Fetherstonhaugh, P.~Slovic, S.~Johnson, and J.~Friedrich.
\newblock Insensitivity to the value of human life: A study of psychophysical
  numbing.
\newblock {\em Journal of Risk and uncertainty}, 14(3):283--300, 1997.

\bibitem{gapminder2016dollar}
{Gapminder Foundation}.
\newblock Dollar street.
\newblock Web page \url{https://www.gapminder.org/dollar-street}. Last visited
  2022-03-22., 2016.

\bibitem{genevsky2013neural}
A.~Genevsky, D.~V{\"a}stfj{\"a}ll, P.~Slovic, and B.~Knutson.
\newblock Neural underpinnings of the identifiable victim effect: Affect shifts
  preferences for giving.
\newblock {\em Journal of Neuroscience}, 33(43):17188--17196, 2013.

\bibitem{givewell2022top}
{GiveWell}.
\newblock Our top charities.
\newblock Web page \url{https://www.givewell.org/charities/top-charities}. Last
  visited 2022-03-22., 2022.

\bibitem{gray2018dark}
C.~M. Gray, Y.~Kou, B.~Battles, J.~Hoggatt, and A.~L. Toombs.
\newblock The dark (patterns) side of ux design.
\newblock In {\em Proceedings of the 2018 CHI conference on human factors in
  computing systems}, pp. 1--14, 2018.

\bibitem{gyroscoping2022building}
{Gyroscoping Games}.
\newblock Building hope: Refugee camp simulator.
\newblock Web page \url{https://gyroscopinggames.com/}. Last visited
  2022-03-23., 2022.

\bibitem{halloran2017emotional}
N.~Halloran.
\newblock Emotional statistics.
\newblock Keynote presentation at Tapestry: The Data Storytelling Conference
  \url{https://www.youtube.com/watch?v=TCqcpL8F99k}., 2017.

\bibitem{halstead2019mental}
J.~Halstead.
\newblock Founders pledge -- mental health executive summary.
\newblock Web article
  \url{https://founderspledge.com/stories/mental-health-report-summary}., 2019.

\bibitem{hamari2014persuasive}
J.~Hamari, J.~Koivisto, and T.~Pakkanen.
\newblock Do persuasive technologies persuade?-a review of empirical studies.
\newblock In {\em International conference on persuasive technology}, pp.
  118--136. Springer, 2014.

\bibitem{harris2015connecting}
J.~Harris.
\newblock Connecting with the dots.
\newblock Web article
  \url{https://source.opennews.org/articles/connecting-dots/}. Last visited
  2022-03-24., 2015.

\bibitem{harris2021can}
S.~Harris, P.~Singer, F.~Minerva, and J.~McMahan.
\newblock Can we talk about scary ideas?
\newblock Making Sense podcast episode \#245
  \url{https://www.samharris.org/podcasts/making-sense-episodes/245-can-talk-scary-ideas},
  2021.

\bibitem{herrera2018building}
F.~Herrera, J.~Bailenson, E.~Weisz, E.~Ogle, and J.~Zaki.
\newblock Building long-term empathy: A large-scale comparison of traditional
  and virtual reality perspective-taking.
\newblock {\em PloS one}, 13(10):e0204494, 2018.

\bibitem{isenberg2018immersive}
P.~Isenberg, B.~Lee, H.~Qu, and M.~Cordeil.
\newblock Immersive visual data stories.
\newblock In {\em Immersive Analytics}, pp. 165--184. Springer, 2018.

\bibitem{ivanov2019cg}
A.~Ivanov, K.~Danyluk, C.~Jacob, and W.~Willett.
\newblock {CG\&A} 2019: A walk among the data: Exploration and anthropomorphism
  in immersive unit visualizations.
\newblock Conference presentation \url{https://vimeo.com/375798028}., 2019.

\bibitem{ivanov2019walk}
A.~Ivanov, K.~Danyluk, C.~Jacob, and W.~Willett.
\newblock A walk among the data.
\newblock {\em IEEE Computer Graphics and Applications}, 39(3):19--28, 2019.

\bibitem{ivanov2018exploration}
A.~Ivanov, K.~T. Danyluk, and W.~Willett.
\newblock Exploration \& anthropomorphism in immersive unit visualizations.
\newblock In {\em Extended Abstracts of the 2018 CHI Conference on Human
  Factors in Computing Systems}, pp. 1--6, 2018.

\bibitem{jansen2015opportunities}
Y.~Jansen, P.~Dragicevic, P.~Isenberg, J.~Alexander, A.~Karnik, J.~Kildal,
  S.~Subramanian, and K.~Hornb{\ae}k.
\newblock Opportunities and challenges for data physicalization.
\newblock In {\em Proc. of the 33rd Annual ACM Conference on Human Factors in
  Computing Systems}, pp. 3227--3236, 2015.

\bibitem{kandaurova2019effects}
M.~Kandaurova and S.~H.~M. Lee.
\newblock The effects of virtual reality (vr) on charitable giving: The role of
  empathy, guilt, responsibility, and social exclusion.
\newblock {\em Journal of Business Research}, 100:571--580, 2019.

\bibitem{kosara2013storytelling}
R.~Kosara and J.~Mackinlay.
\newblock Storytelling: The next step for visualization.
\newblock {\em Computer}, 46(5):44--50, 2013.

\bibitem{liberman2007psychological}
N.~Liberman, Y.~Trope, and E.~Stephan.
\newblock Psychological distance.
\newblock 2007.

\bibitem{luckner2008prediction}
S.~Luckner.
\newblock Prediction markets: Fundamentals, key design elements, and
  applications.
\newblock {\em BLED 2008 proceedings}, p.~27, 2008.

\bibitem{lupi2017data}
G.~Lupi.
\newblock Data humanism: the revolution will be visualized.
\newblock Web article
  \url{http://giorgialupi.com/data-humanism-my-manifesto-for-a-new-data-wold}.
  Last visited 2022-03-30., 2017.

\bibitem{macaskill2020can}
W.~MacAskill and S.~Harris.
\newblock Doing good.
\newblock Making Sense podcast episode \#228
  \url{https://www.samharris.org/podcasts/making-sense-episodes/228-doing-good},
  2020.

\bibitem{madsen2015interview}
L.~Madsen and P.~Dragicevic.
\newblock Interview with loren madsen: The birth of data sculpture.
\newblock Web article \url{http://dataphys.org/list/loren-madsen-interview/}.,
  2013.

\bibitem{marriott2018immersive}
K.~Marriott, F.~Schreiber, T.~Dwyer, K.~Klein, N.~H. Riche, T.~Itoh,
  W.~Stuerzlinger, and B.~H. Thomas.
\newblock {\em Immersive analytics}, vol. 11190.
\newblock Springer, 2018.

\bibitem{mars2019known}
R.~Mars and J.~Rosenberg.
\newblock The known unknown.
\newblock The 99\% Invisible Podcast episode \#344
  \url{https://99percentinvisible.org/episode/the-known-unknown/}, 2019.

\bibitem{mars2019sound}
R.~Mars, J.~Schlesinger, J.~Beckerman, and Y.~Sen.
\newblock Sound and health: Hospitals.
\newblock The 99\% Invisible Podcast
  \url{https://99percentinvisible.org/episode/sound-and-health-hospitals/},
  2019.

\bibitem{martingano2021virtual}
A.~J. Martingano, F.~Hererra, and S.~Konrath.
\newblock Virtual reality improves emotional but not cognitive empathy: a
  meta-analysis.
\newblock 2021.

\bibitem{meikle2013poverty}
G.~Meikle.
\newblock Poverty porn: is sensationalism justified if it helps those in need?
\newblock Web article
  \url{https://www.theguardian.com/global-development-professionals-network/2013/jul/05/poverty-porn-development-reporting-fistula}.
  Last visited 2022-03-25., 2013.

\bibitem{menzel1994material}
P.~Menzel and C.~C. Mann.
\newblock {\em Material world: a global family portrait}.
\newblock Univ of California Press, 1994.

\bibitem{milk2015how}
C.~Milk.
\newblock How virtual reality can create the ultimate empathy machine.
\newblock {TED} Talk
  \url{https://www.ted.com/talks/chris_milk_how_virtual_reality_can_create_the_ultimate_empathy_machine}.,
  2015.

\bibitem{moere2007towards}
A.~V. Moere.
\newblock Towards designing persuasive ambient visualization.
\newblock In {\em Issues in the Design \& Evaluation of Ambient Information
  Systems Workshop}, pp. 48--52. Citeseer, 2007.

\bibitem{morais2020list}
L.~Morais.
\newblock List of anthropographics -- a collection of data visualizations about
  people.
\newblock Web page \url{https://luizaugustomm.github.io/anthropographics/}.
  Last visited 2022-03-22., 2018.

\bibitem{morais2022exploring}
L.~Morais, N.~Andrade, and D.~Sousa.
\newblock {Exploring How Visualization Design and Situatedness Evoke Compassion
  in the Wild}.
\newblock In {\em {Eurographics Conference on Visualization (EuroVis) 2022}}.
  Rome, Italy, June 2022.

\bibitem{morais2020showing}
L.~Morais, Y.~Jansen, N.~Andrade, and P.~Dragicevic.
\newblock Showing data about people: A design space of anthropographics.
\newblock {\em IEEE Transactions on Visualization and Computer Graphics}, 2020.

\bibitem{morais2021can}
L.~Morais, Y.~Jansen, N.~Andrade, and P.~Dragicevic.
\newblock Can anthropographics promote prosociality? a review and large-sample
  study.
\newblock In {\em Proceedings of the 2021 CHI Conference on Human Factors in
  Computing Systems}, pp. 1--18, 2021.

\bibitem{noetel2020we}
M.~Noetel, P.~Slattery, A.~K. Saeri, J.~Lee, T.~Houlden, N.~Farr, R.~Gelber,
  J.~Stone, L.~Huuskes, S.~Timmons10, et~al.
\newblock How do we get people to donate more to charity? an overview of
  reviews.
\newblock 2020.

\bibitem{numinous2016that}
{Numinous Games}.
\newblock That dragon, cancer.
\newblock Web page \url{http://www.thatdragoncancer.com/}. Last visited
  2022-03-23., 2016.

\bibitem{ord2013moral}
T.~Ord.
\newblock The moral imperative toward cost-effectiveness in global health.
\newblock {\em Center for Global Development}, pp. 1--8, 2013.

\bibitem{orts2016holoportation}
S.~Orts-Escolano, C.~Rhemann, S.~Fanello, W.~Chang, A.~Kowdle, Y.~Degtyarev,
  D.~Kim, P.~L. Davidson, S.~Khamis, et~al.
\newblock Holoportation: Virtual 3d teleportation in real-time.
\newblock In {\em Proc. 29th annual symposium on user interface software and
  technology}, pp. 741--754, 2016.

\bibitem{peck2013putting}
T.~C. Peck, S.~Seinfeld, S.~M. Aglioti, and M.~Slater.
\newblock Putting yourself in the skin of a black avatar reduces implicit
  racial bias.
\newblock {\em Consciousness and cognition}, 22(3):779--787, 2013.

\bibitem{phelps2021killing}
L.~C.~W. Phelps.
\newblock {\em On Killing Remotely: The Psychology of Killing with Drones}.
\newblock Hachette UK, 2021.

\bibitem{pimentel2021peril}
D.~Pimentel.
\newblock The peril and potential of {XR}-based interactions with wildlife.
\newblock In {\em Extended Abstracts of the 2021 CHI Conference on Human
  Factors in Computing Systems}, pp. 1--9, 2021.

\bibitem{pousman2006taxonomy}
Z.~Pousman and J.~Stasko.
\newblock A taxonomy of ambient information systems: four patterns of design.
\newblock In {\em Proceedings of the working conference on Advanced visual
  interfaces}, pp. 67--74, 2006.

\bibitem{ricci2012social}
F.~Ricci.
\newblock Social implications of malaria and their relationships with poverty.
\newblock {\em Med. j. of hematology and infectious diseases}, 4(1), 2012.

\bibitem{riche2018data}
N.~H. Riche, C.~Hurter, N.~Diakopoulos, and S.~Carpendale.
\newblock {\em Data-driven storytelling}.
\newblock CRC Press, 2018.

\bibitem{roseway2015biocrystal}
A.~Roseway, Y.~Lutchyn, P.~Johns, E.~Mynatt, and M.~Czerwinski.
\newblock Biocrystal: An ambient tool for emotion and communication.
\newblock {\em Intl J. of Mobile Human Computer Interaction (IJMHCI)},
  7(3):20--41, 2015.

\bibitem{rosling2010magic}
H.~Rosling.
\newblock The magic washing machine.
\newblock {TEDWomen} Talk
  \url{https://www.ted.com/talks/hans_rosling_the_magic_washing_machine}.,
  2010.

\bibitem{sauve2020econundrum}
K.~Sauv{\'e}, S.~Bakker, and S.~Houben.
\newblock Econundrum: Visualizing the climate impact of dietary choice through
  a shared data sculpture.
\newblock In {\em Proc. DIS}, pp. 1287--1300, 2020.

\bibitem{segel2010narrative}
E.~Segel and J.~Heer.
\newblock Narrative visualization: Telling stories with data.
\newblock {\em IEEE transactions on visualization and computer graphics},
  16(6):1139--1148, 2010.

\bibitem{serino2016virtual}
S.~Serino, E.~Pedroli, A.~Keizer, S.~Triberti, A.~Dakanalis, F.~Pallavicini,
  A.~Chirico, and G.~Riva.
\newblock Virtual reality body swapping: a tool for modifying the allocentric
  memory of the body.
\newblock {\em Cyberpsychology, Behavior, and Social Networking},
  19(2):127--133, 2016.

\bibitem{shneiderman2013improving}
B.~Shneiderman, C.~Plaisant, and B.~W. Hesse.
\newblock Improving healthcare with interactive visualization.
\newblock {\em Computer}, 46(5):58--66, 2013.

\bibitem{sillesen2015journalism}
L.~B. Sillesen, C.~Ip, and D.~Uberti.
\newblock Journalism and the power of emotions.
\newblock {\em Columbia Journalism Review}, 54(1):15, 2015.

\bibitem{singer1972morality}
P.~Singer.
\newblock Famine, affluence, and morality.
\newblock {\em Philosophy and Public Affairs}, 1(3), 1972.

\bibitem{singer2009life}
P.~Singer.
\newblock {\em The Life You Can Save: How to Do Your Part to End World
  Poverty}.
\newblock Random House, 2009.

\bibitem{singer2011expanding}
P.~Singer.
\newblock {\em The expanding circle: Ethics, evolution, and moral progress}.
\newblock Princeton University Press, 2011.

\bibitem{singer2013why}
P.~Singer.
\newblock The why and how of effective altruism.
\newblock {TED} Talk (be warned that the talk starts with a shocking video
  scene)
  \url{https://www.ted.com/talks/peter_singer_the_why_and_how_of_effective_altruism}.,
  2013.

\bibitem{skog2003between}
T.~Skog, S.~Ljungblad, and L.~E. Holmquist.
\newblock Between aesthetics and utility: designing ambient information
  visualizations.
\newblock In {\em IEEE Symposium on Information Visualization 2003 (IEEE Cat.
  No. 03TH8714)}, pp. 233--240. IEEE, 2003.

\bibitem{soryl2021case}
A.~A. Soryl, A.~J. Moore, P.~J. Seddon, and M.~R. King.
\newblock The case for welfare biology.
\newblock {\em Journal of Agricultural and Environmental Ethics}, 34(2):1--25,
  2021.

\bibitem{pixel2018bury}
{The Pixel Hunt}, {Figs}, and {ARTE France}.
\newblock Bury me, my love.
\newblock Web page \url{https://burymemylove.arte.tv/contact/}. Last visited
  2022-03-23., 2017.

\bibitem{thomas2018situated}
B.~H. Thomas, G.~F. Welch, P.~Dragicevic, N.~Elmqvist, P.~Irani, Y.~Jansen,
  D.~Schmalstieg, A.~Tabard, N.~A. ElSayed, et~al.
\newblock Situated analytics.
\newblock In {\em Immersive analytics}, vol. 11190, pp. 185--220. 2018.

\bibitem{tomasik2018how}
B.~Tomasik.
\newblock How much direct suffering is caused by various animal foods?
\newblock Web article
  \url{https://reducing-suffering.org/how-much-direct-suffering-is-caused-by-various-animal-foods/}.
  Last visited 2022-03-23., 2007--2018.

\bibitem{tomasik2019brain}
B.~Tomasik.
\newblock Is brain size morally relevant?
\newblock Web article
  \url{https://reducing-suffering.org/is-brain-size-morally-relevant/}. Last
  visited 2022-03-23., 2013--2019.

\bibitem{tomasik2018trc}
B.~Tomasik, D.~McKee, and A.~Gardner.
\newblock Brian tomasik on wild animal suffering.
\newblock The Reality Check Podcast episode \#490: Mashup
  \url{http://www.trcpodcast.com/trc-490-mashup-brian-tomasik-on-wild-animal-suffering/},
  2018.

\bibitem{verge2014using}
T.~Verge.
\newblock Using the oculus rift to enter the body of another.
\newblock Video \url{https://www.youtube.com/watch?v=dOSJETowuik}., 2014.

\bibitem{VFSG}
{VFSG}.
\newblock Viz for social good.
\newblock Web page \url{https://www.vizforsocialgood.com/}. Last visited
  2022-03-22., 2022.

\bibitem{vinding2020suffering}
M.~Vinding.
\newblock Suffering-focused ethics: Defense and implications.
\newblock {\em Independently published, May}, 2020.

\bibitem{werner2008united}
J.~Werner, R.~Wettach, and E.~Hornecker.
\newblock United-pulse: feeling your partner's pulse.
\newblock In {\em Proc. 10th intl. conference on Human computer interaction
  with mobile devices and services}, pp. 535--538, 2008.

\bibitem{willett2021superpowers}
W.~Willett, B.~A. Aseniero, S.~Carpendale, P.~Dragicevic, Y.~Jansen,
  L.~Oehlberg, and P.~Isenberg.
\newblock Superpowers as inspiration for visualization.
\newblock {\em IEEE TVCG}, 2021, 2021.

\bibitem{willett2016embedded}
W.~Willett, Y.~Jansen, and P.~Dragicevic.
\newblock Embedded data representations.
\newblock {\em IEEE transactions on visualization and computer graphics},
  23(1):461--470, 2016.

\bibitem{wisneski1998ambient}
C.~Wisneski, H.~Ishii, A.~Dahley, M.~Gorbet, S.~Brave, B.~Ullmer, and P.~Yarin.
\newblock Ambient displays: Turning architectural space into an interface
  between people and digital information.
\newblock In {\em International Workshop on Cooperative Buildings}, pp. 22--32.
  Springer, 1998.

\bibitem{world2022vitamin}
{World Health Organization}.
\newblock Vitamin {A} deficiency.
\newblock Web article
  \url{https://www.who.int/data/nutrition/nlis/info/vitamin-a-deficiency}. Last
  visited 2022-03-22., 2022.

\bibitem{zhang2021mapping}
Y.~Zhang, Y.~Sun, L.~Padilla, S.~Barua, E.~Bertini, and A.~G. Parker.
\newblock Mapping the landscape of covid-19 crisis visualizations.
\newblock In {\em Proc. 2021 CHI Conf. on Human Factors in Computing Systems},
  pp. 1--23, 2021.

\end{thebibliography}
\end{document}